\documentclass[a4paper,fleqn,usenatbib]{mnras}

\usepackage{newtxtext,newtxmath}
\usepackage[T1]{fontenc}
\usepackage{ae,aecompl}

\usepackage{graphicx}	% Including figure files
\usepackage{amsmath}	% Advanced maths commands
\usepackage{amssymb}	% Extra maths symbols

\title[A methodology to select galaxies just after the quenching of star formation]{A methodology to select galaxies just after the quenching of star formation}

\author[A. Citro et al.]{
Annalisa Citro,$^{1,2}$\thanks{E-mail: annalisa.citro@unibo.it}
Lucia Pozzetti,$^{2}$
Salvatore Quai,$^{1, 2}$
Michele Moresco,$^{1, 2}$
\newauthor~Livia Vallini$^{1, 2, 4}$
and Andrea Cimatti$^{1, 3}$
\\
$^{1}$Dipartimento di Fisica e Astronomia, Universit\`{a} di Bologna, Via Gobetti 93/2, I-40129, Bologna, Italy\\
$^{2}$INAF - Osservatorio Astronomico di Bologna, Via Gobetti 93/3, I-40129, Bologna, Italy\\
$^{3}$INAF - Osservatorio Astrofisico di Arcetri, Largo E. Fermi 5, I-50125, Firenze, Italy\\
$^{4}$Nordita, KTH Royal Institute of Technology and Stockholm University, Roslagstullsbacken 23, SE-106 91 Stockholm, Sweden
}

\date{Accepted XXX. Received YYY; in original form ZZZ}

\pubyear{2015}

\begin{document}
\label{firstpage}
\pagerange{\pageref{firstpage}--\pageref{lastpage}}
\maketitle

\begin{abstract}
We propose a new methodology aimed at finding star-forming galaxies in the phase which immediately follows the star-formation (SF) quenching, based on the use of high- to low-ionization emission line ratios. These ratios rapidly disappear after the SF halt, due to the softening of the UV ionizing radiation. We focus on [O III] $\lambda$5007/H$\alpha$ and [Ne III] $\lambda$3869/[O II] $\lambda$3727, studying them with simulations obtained with the CLOUDY photoionization code.\textcolor{black}{ If a sharp quenching is assumed, we find that the two ratios are very sensitive tracers as they drop by a factor $\sim$ 10 within $\sim$ 10 Myr from the interruption of the SF; instead, if a smoother and slower SF decline is assumed (i.e. an exponentially declining star-formation history with $e$-folding time $\tau=$ 200 Myr), they decrease by a factor $\sim$ 2 within $\sim$ 80 Myr.} %We find that, if a sharp quenching is assumed, the two ratios trace it on very short time-scales, dropping by a factor $\gtrsim$ 10 within $\sim$ 10 Myr from the SF halt; if a smoother and slower SF decline is assumed (i.e. an exponentially declining star-formation history with $e$-folding time $\tau=$ 200 Myr), 
%they instead decrease by a factor $\sim$ 2 within $\sim$ 80 Myr.} 
%the time interval needed by the star-forming region to become quiescent (i.e. $\sim$ 80 Myr).} 
%Moreover, the [O III] $\lambda$5007/H$\beta$ ratio is also proved to be an useful dust-free, quenching diagnostic. 
We mitigate the ionization -- metallicity degeneracy affecting our methodology using pairs of emission line ratios separately related to metallicity and ionization, adopting the [N II] $\lambda$6584/[O II] $\lambda$3727 ratio as metallicity diagnostic. Using a Sloan Digital Sky Survey galaxy sample, \textcolor{black}{we identify 10 examples among the most extreme quenching candidates within the [O III] $\lambda$5007/H$\alpha$ vs. [N II] $\lambda$6584/[O II] $\lambda$3727 plane, characterized by low [O III] $\lambda$5007/H$\alpha$}, \textcolor{black}{faint [Ne III] $\lambda$3869, and by blue dust-corrected spectra and $(u-r)$ colours, as expected if the SF quenching has occurred in the very recent past.} Our results also suggest that the observed fractions of quenching candidates can be used to constrain the quenching mechanism at work \textcolor{black}{and its time-scales}.

\end{abstract}


%%%%%%%%%%%%%%%%%%%%%%%%%%%%%%%%%%%%%%%%%%%%%%%%%%

%%%%%%%%%%%%%%%%% BODY OF PAPER %%%%%%%%%%%%%%%%%%

\section{Introduction}

One of the key question in the current studies of galaxy evolution is to understand the origin of the dichotomy which segregates galaxies into the blue cloud (i.e. late-type, star-forming and disc dominated galaxies) and the red sequence (i.e. early-type, quiescent and bulge dominated galaxies). For $M>10^{10} M_{\sun}$, this bimodality is clearly visible both at $z\sim0$ (\citealp{Strateva+2001}; \citealp{Blanton+2003c}; \citealp{Kauffmann+2003a}; \citealp{Baldry+2004}) and at higher redshifts (\citealp{Bell+2004a}, \citealp{Willmer+2006}, \citealp{Cassata+2008}, \citealp{Brammer+2009}, \citealp{Whitaker+2011}, \citealp{Wuyts+2011}, \citealp{Bell+2012}, \citealp{Cheung+2012}, \citealp{Muzzin+2013}) and concerns both galaxy colours and structure (\citealp{McIntosh+2014}).\\
There is a general consensus that the main process underlying the galaxy bimodality is the build-up of the red sequence population at $M<10^{11} M_{\sun}$, driven by the quenching of the star formation and by a morphological transformation, which turns late-type into early-type morphologies (e.g. \citealp{Drory+2004}, \citealp{Faber+2005}, \citealp{Pozzetti+2010}). \\
What is still unconstrained are the mechanisms able to modify both star-formation and morphology and their corresponding time-scales. Several hypothesis have been proposed to explain the quenching of the star formation in blue galaxies, such as gas stripping (e.g. \citealp{Gunn&Gott1972}), morphological or gravitational quenching (\citealp{Martig+2009}, \citealp{Genzel+2014}), shock heating of infalling cold gas by the hot halo (\citealp{Dekel&Birnboim2006}), or an exhaustion of the gas supply (e.g. \citealp{Larson+1980}). Moreover, in massive galaxies, the role of AGNs in influencing galaxy evolution and quenching star formation is supported by several observations (\citealp{Hopkins+2005}; \citealp{Kaviraj+2007b};  \citealp{Diamond-Stanic+2012}; \citealp{Fabian2012} and references therein, \citealp{Cimatti+2013}, \citealp{Cicone+2014}, \citealp{ForsterSchreiber+2014}), and corroborated by the theoretical results obtained combining N-body simulations of dark matter halos evolution (\citealp{Springel+2005}, \citealp{Boylan-Kolchin+2009}) with semi-analytic models for galaxy formation (\citealp{White&Frenk1991}, \citealp{Springel+2005}, \citealp{Lu+2011}, \citealp{Benson2012}). However, other models are capable to form rapidly quiescent galaxies without invoking the AGN feedback (e.g. \citealp{Naab+2006}, \citeyear{Naab+2009}; \citealp{Khochfar&Silk2006}, \citealp{Johansson+2012}). Stellar or supernova (SN) feedback is most likely channel for the star formation (SF) quenching in low-mass galaxies (e.g. $<10^{10} M_{\sun}$; \citealp{Kaviraj+2007b}). 
Several mechanisms have also been invoked to explain the morphological transformation. Numerical simulations have shown that major merging can give rise to elliptical and S0 galaxies (\citealp{Bekki+1998}) and that also minor merging can play an important role in spheroid and bulge growth (\citealp{Bournaud+2007}; \citealp{Naab+2007}). From an observational point of view, evidences that  the morphological transformation can be also induced by environmental mechanisms (\citealp{Larson+1980}; \citealp{Bekki+2002}, \citealp{Farouki&Shapiro1981}; \citealp{Moore+1999}, \citealp{Quilis+2000}) or by the secular growth of pseudo-bulges (\citealp{Courteau+1996}; \citealp{Norman+1996}; \citealp{MacArthur+2003}; \citealp{Kormendy&Kennicutt2004}; \citealp{Debattista+2006}) have been found.\\
Having intermediate colours between the blue cloud and the red sequence, galaxies populating the so-called 'green-valley' (e.g. \citealp{Salim2014}, \citealp{Schawinski+2014}) are generally considered as the transiting objects par excellence (\citealp{Martin+2007}, \citealp{Mendel+2013}, \citealp{Salim2014}, \citealp{Schawinski+2014}). 
Among these, the most interesting population certainly consists of those galaxies which have just entered the quenching phase (within a few Myr). Although hampered by the short duration of the quenching process, the search for galaxies in this critical phase of evolution has been carried on by several authors in the past decades. Galaxies characterized by both a tidally disturbed morphology and intermediate colours (e.g. \citealp{Schweizer&Seitzer1992}; \citealp{Tal+2009}) or low-level of recent SF (\citealp{Kaviraj2010}), young elliptical galaxies (\citealp{Sanders+1988}; \citealp{Genzel+2001}; \citealp{Dasyra+2006}) and very recent post-merger remnants with strong morphological disturbances (\citealp{Hibbard+1996}; \citealp{Rothberg&Joseph2004}; \citealp{Carpineti+2012}) have been considered as valid 'recent time' quenching candidates. Moreover, many attempts aimed at spectroscopically identifying quenching galaxies come from the investigations of the post-starburst (E+A or K+A) galaxies UV and optical spectra, whose strong Balmer absorption lines and missing [O II] $\lambda$3727 ([O II], hereafter) and H$\alpha$ emission lines (\citealp{Couch&Sharples1987}, \citealp{Quintero+2004}, \citealp{Poggianti+2004}; \citealp{Balogh+2011}; \citealp{Muzzin+2012}; \citealp{Mok+2013}; \citealp{Wu+2014}) have been interpreted as signs of a recent halt of the SF (\citealp{Dressler&Gunn1983}; \citealp{Zabludoff+1996}; \citealp{Quintero+2004}, \citealp{Poggianti+2008}, \citealp{Wild+2009}).
The scarcity of galaxies that are in the transition phase suggests that, whatever mechanism may be responsable of the star formation shut-off, it has to happen on short time-scales (\citealp{Tinker+2010}, \citealp{Salim2014}). \textcolor{black}{The relatively short duration of the quenching process is also suggested by the surprising identification of a significant number of galaxies that look already quiescent at $z \sim 4 - 5$, when the Universe was only $\sim$ 1 -- 1.5 Gyr old. %which have stellar masses up to $M \sim 10^{11} M_{\bigodot}$ and 
(e.g. \citealp{Mobasher+2005}, \citealp{Wiklind+2008}, \citealp{Juarez+2009}, \citealp{Brammer+2011}, \citealp{Marsan+2015}, \citealp{Citro+2016})}. However, a general and coherent picture concerning the SF quenching is still lacking. Very recently, evidences that the quenching of the star formation could be a separated process with respect to the morphological transformation have come from the photometric and spectroscopic investigations of passive spiral galaxies (\citealp{Fraser-McKelvie+2016}).\\ Further uncertainties concern the typical time-scales of the quenching process, which are far to be constrained. For instance, galaxies with early-type morphologies seem to be characterized by shorter quenching time-scales, % \textcolor{red}{which can also be compatible with an instantaneus SF shut-off}, 
with respect to galaxies with late-type morphologies (\citealp{Schawinski+2014}), and the time-scales derived observationally are generally shorter than the ones derived using N-body simulations and semi-analytic models (\citealp{Muzzin+2012}; \citealp{Wetzel+2013}; \citealp{Mok+2013}, \citealp{Taranu+2014}).\\
Emission line ratios have been proved to be powerful tools to constrain the ionization state and the properties of galaxies both at $z\sim0$ and at higher redshift (e.g. \citealp{Dopita+2000}, \citealp{Kewley+2001}, \citealp{Dopita+2006}, \citealp{Levesque+2010}, \citealp{Kewley+2013}, \citealp{Kashino+2016}), but this kind of studies are always hampered by the degeneracies affecting spectra and emission lines, which in many cases make it difficult to disentangle what is the origin of the emission line ratios intensity (e.g. \citealp{Dopita+2006}). However, a general view on how the emission line ratios can help in tracing the quenching phase is lacking, especially when the very early epochs after the quenching are involved.\\
\textcolor{black}{In this work, we propose a methodology aimed at identifying galaxies in the phase which immediately follows the quenching of the SF, based on the search for galaxies with weak high-ionization and strong low-ionization emission lines in their spectra. In particular, we adopt the [O III] $\lambda$5007 ([O III], hereafter) and [Ne III] $\lambda$3869 ([Ne III], hereafter) high-ionization lines, and the H$\alpha$ and [O II] $\lambda3727$ ([O II], hereafter) low-ionization lines. [O III] and [Ne III] probe the presence of short-lived and very massive O stars able to provide photons hard enough to doubly ionize oxygen and neon; H$\alpha$ and [O II], being excited by softer photons, probe the presence of colder O stars and B stars.}\\
We simulate the emission line ratios using the photoionization code CLOUDY (\citealp{Ferland1998}), \textcolor{black}{assuming short time-scales for the quenching}. We also combine the emission line ratios with metallicity diagnostics, in order to mitigate the ionization -- metallicity degeneracy.

\section{The proposed approach}

\label{sect:approach}
We propose a methodology able to identify galaxies in the phase just after the quenching of the star formation, based on emission line ratios. % \textcolor{black}{We mostly rely on the assumption of a sharp quenching of the SF, according to recent studies which argue that the quenching process can occur on very short time-scales of less than $100-200$ Myr (e.g. \citealp{Schawinski+2014}, \citealp{Ciesla+2016}, \citealp{Schaefer+2017}). However, during our analysis, we also illustrate the results obtained if a smoother decline of the SF is assumed.}\\
If the star formation is quenched in a star-forming region, the number of UV hydrogen ionizing photons (i.e. $\lambda<912$ \AA) provided by the central ionizing source declines. In particular, since massive O stars are the first to disappear, the harder UV photons are the first to drop, leading to a softening of the UV ionizing spectrum. As a consequence, high-ionization lines, which can be excited only by the hardest ionizing photons, rapidly disappear; \textcolor{black}{low-ionization lines, which can be produced also by colder O stars and B stars, remain strong at later times.} \\\textcolor{black}{This allows us to devise a new methodology for identifying quenching galaxies. We search for galaxy spectra characterized by very faint high-ionization lines, \textcolor{black}{like [O III] and [Ne III], but still strong low-ionization lines, like H$\alpha$ and [O II]}. We put into practice our methodology by looking for galaxy spectra with high signal-to-noise (S/N) H$\alpha$ emission but very low values of the [O III]/H$\alpha$  and the [Ne III]/[O II] emission line ratios. These lines are observable in a wide range of redshifts (e.g. 0 < z < 2) with optical and near infrared spectroscopy.} \textcolor{black}{In particular, [Ne III]/[O II], despite involving a very faint emission line like [Ne III] $\lambda$3869 ([Ne III], hereafter), has the advantage of being basically unaffected by dust extinction, including emission lines which are very close in wavelength. We also investigated the possibility to use a line ratio less affected by dust extinction correction such as the [O III] $\lambda$5007/H$\beta$ ([O III]/H$\beta$, hereafter), which has also the advantage of involving stronger emission lines than [Ne III]/[O II].}\\
\textcolor{black}{\textcolor{black}{In the following, we assume a sharp quenching of the SF as extreme case. Moreover, we illustrate the results obtained if a more realistic, smoother  but still short decline of the SF is adopted.
This is in agreement with recent studies arguing that quenching processes occurring on 100 -- 200 Myr time-scales can be modelled by a sudden interruption of the SF (e.g. \citealp{Schawinski+2014}, \citealp{Ciesla+2016}, \citealp{Schaefer+2017}).}}\\
  
\section{Modelling the quenching phase}

\label{sec:photoionmodel}
\textcolor{black}{In order to investigate the behaviour of the proposed emission line ratios during the quenching phase, we simulate \textcolor{black}{star-forming regions} until their quiescent phase. Here we describe the main ingredients of our photoionization models.}\\
We consider \textcolor{black}{a star-forming region} as formed by a central source of energy, with a given spectral energy distribution (SED) and intensity, surrounded by a spherical cloud. We simulate its final spectrum by means of the photoionization code CLOUDY (Version 13.03, \citealp{Ferland1998}, \citealp{Ferland+2013}), adopting a plane-parallel geometry (i.e. the simplest geometry allowed by the code), in which the thickness of the photoionized nebula is very small compared to the distance to the photoionizing source. \\
\textcolor{black}{The shape} of the ionizing source is simulated by means of different stellar synthetic spectra, such as the \citealp{Leitherer+1999} (Starburst99) and  \citealp{Bruzual&Charlot2003} (BC03) models. \\ 
The adopted Starburst99 synthetic spectra are simple stellar populations (SSPs) with a fixed mass of $M=10^6~M_{\sun}$, metallicities Z = 0.004, 0.008, 0.02, 0.04 and a Salpeter initial mass function (IMF) (with $m_{\rm{low}}$=1 and $m_{\rm{up}}$=100 M$_{\sun}$). They are computed using Lejeune -- Schmutz stellar atmospheres (\citealp{Lejeune+1997}; \citealp{Schmutz+1992}) and Geneva-HIGH 1994 evolutionary tracks (\citealp{Meynet+1994}, \citealp{Leitherer+1999}).
This set of ingredients is in agreement with the ones generally used in the literature (e.g. \citealp{Levesque+2010}, \citealp{Kewley+2001}, \citealp{Dopita+2006}). \\
For BC03 synthetic spectra, we adopt SSPs with metallicities Z = 0.004, 0.008, 0.02, \textcolor{black}{0.04. In particular, since the default BC03 highest metallicity is Z = 0.05, we interpolate the metallicities to create BC03 models with Z = 0.04, in order to be consistent with the Starburst99 results.} These models are normalized to $M=1~M_{\sun}$ and assume a Chabrier IMF with $m_{\rm{low}}$=0.1 and $m_{\rm{up}}$=100 $M_{\sun}$. \textcolor{black}{The slope of these two IMFs differs at $M < 1~M_{\sun}$, but  we expect this difference not to influence the shape of the final SED, since very low mass stars contribute mostly to the mass of a stellar population rather than to its UV spectral properties. \textcolor{black}{Moreover, since we are interested in emission line ratios, we can neglect the different mass normalization of the two models.}}\\

\textcolor{black}{The intensity of the ionizing source is parametrized by the adimensional ionization parameter $U$, which is defined as the ratio between the mean intensity of the radiation field and the density of the ionized gas.} \textcolor{black}{In the plane parallel case, $U$ can be written as (\citealp{Tielens})}:
\begin{equation}
U=F_{\rm{0}}/n_{\rm{H}}c~~,
\label{eq:U}
\end{equation}
where $n_{\rm{H}}$ is the hydrogen number density of the photoionized gas, $c$ is the speed of light and $F_{\rm{0}}$ \textcolor{black}{is the flux of the UV ionizing photons ($\lambda$ < 912 \AA ) \textcolor{black}{striking} the photoionized cloud. $F_{\rm{0}}$ \textcolor{black}{is proportional to the number of UV hydrogen ionizing photons $Q(H)$, which in turns} depends on the stellar metallicity, the stellar mass, the star-formation rate, the age and the IMF of the ionizing central source. \textcolor{black}{For instance, higher masses, higher star formation rates (SFRs), younger ages or top-heavy IMFs, which all imply an higher number of massive stars, lead $Q(H)$ to increase.} Moreover, $F_{\rm{0}}$ depends on the proximity of the central stars to the photoionized nebula. As explained in Sect.  \ref{sect:approach}, when a galaxy quenches its star formation, $Q(H)$ declines due to the aging and the softening of the ionizing SED. As a consequence, $F_{\rm{0}}$ decreases and so does the ionization parameter.}\\
\textcolor{black}{Throughout this work, we adopt two kind of models, accounting for the decrease of the ionization parameter and the decline of the number of ionizing photons due to the quenching process.}\\

\begin{figure}%[t!]
    \resizebox{\hsize}{!}{\includegraphics{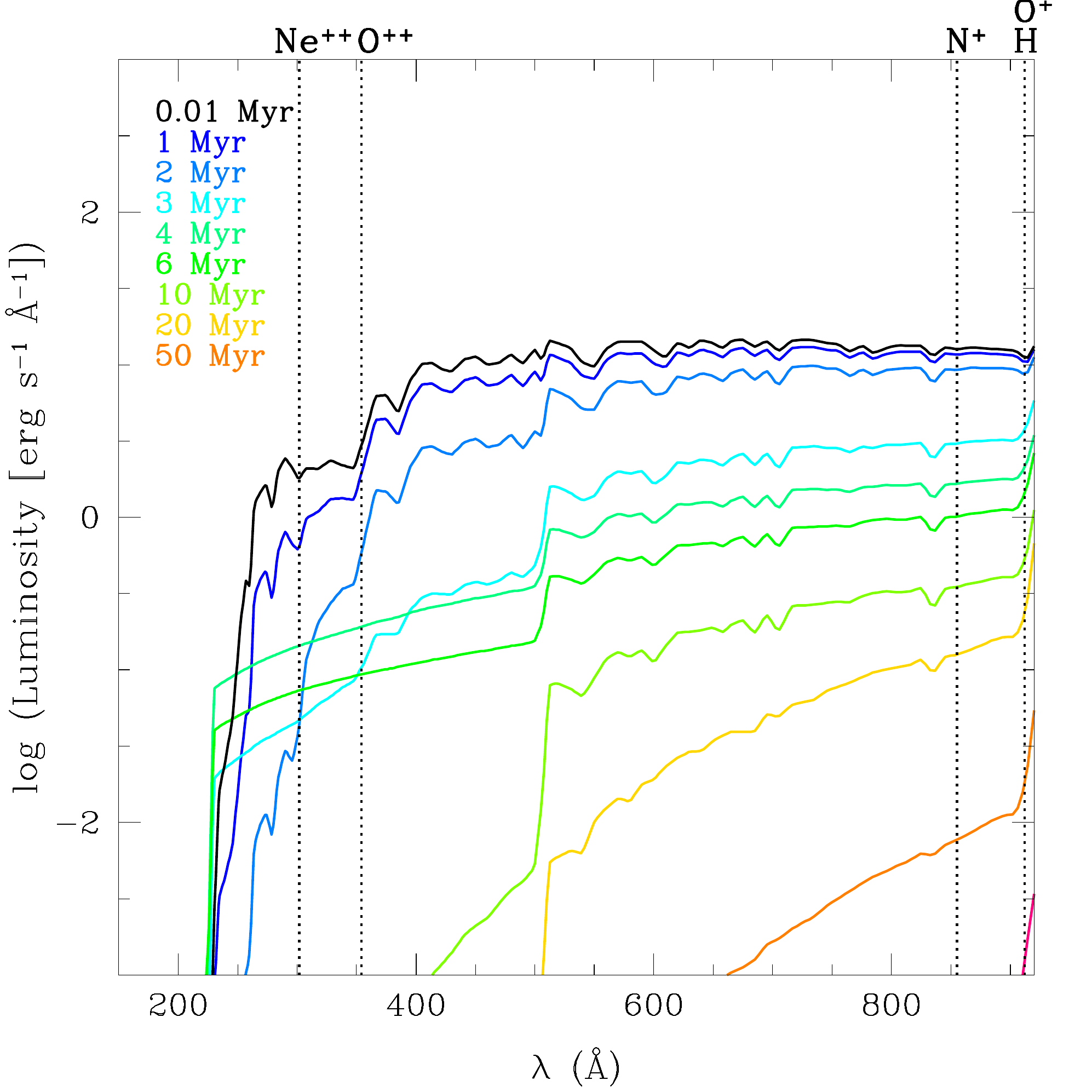}}
          \caption{Time evolution of a Starburst99 SSP SED with solar metallicity (spectra get older from blue to orange, as reported in the top left of the figure). Black dotted vertical lines indicate the wavelengths corresponding to the ionization energies of the emission lines analysed in this work, as indicated.
          }
    \label{fig:Stb99_gen94}
    \end{figure}

\begin{figure}%[t!]
    \resizebox{\hsize}{!}{\includegraphics{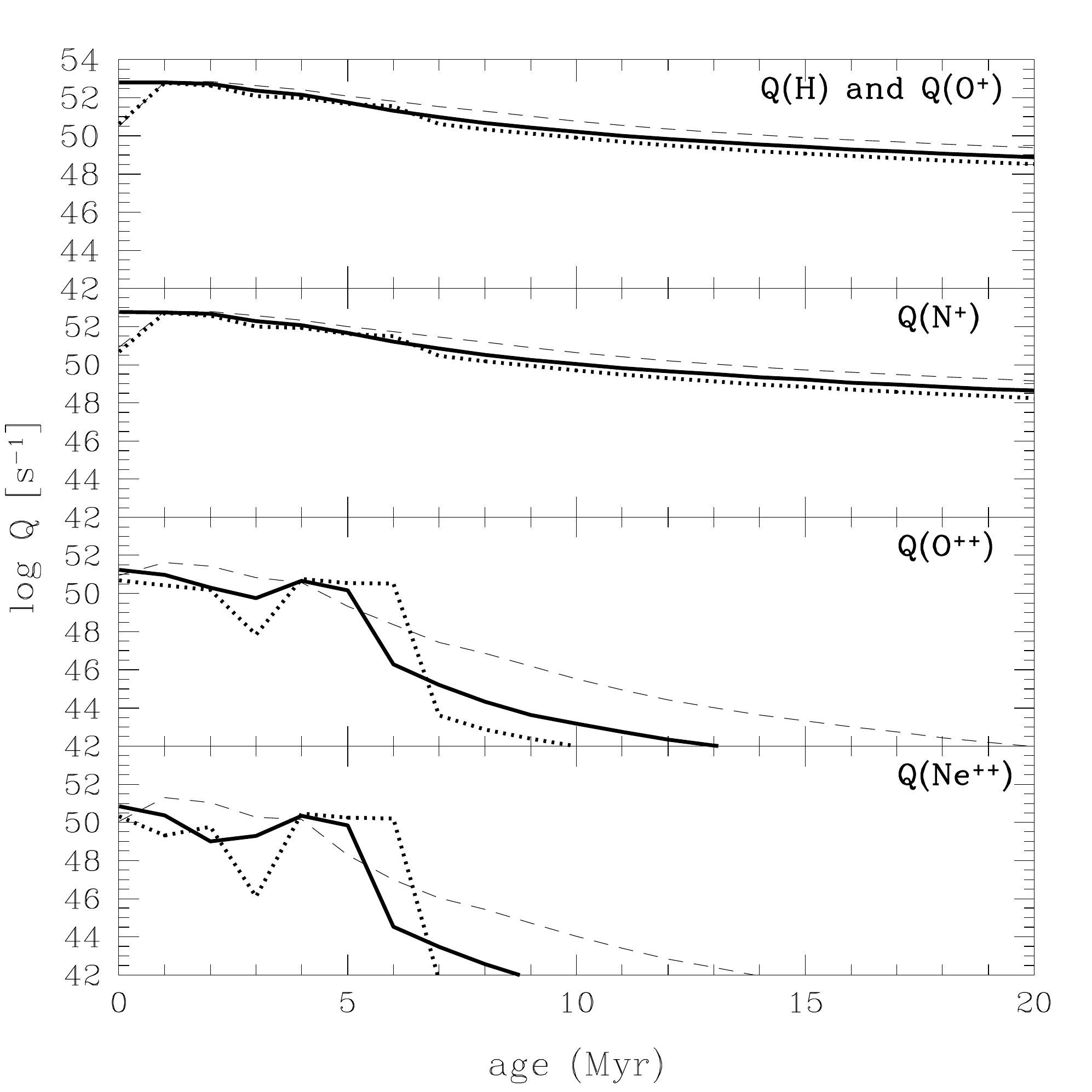}}
     \caption{Time evolution of the number of photons able to ionize H and O$^+$, N$^+$, O$^{++}$ and Ne$^{++}$, from top to bottom.  \textcolor{black}{Note that H and O$^+$ have the same ionization potential (i.e. 13.6 eV) and thus are illustrated within the same panel.} Curves are relative to a Starburst99 SSP with log$(M/M_{\sun})=10^{6}$ and three different metallicities (Z = 0.004, dashed; Z = 0.02, solid; Z = 0.04, dotted).}
    \label{fig:nphot}
    \end{figure}

\begin{enumerate}

\item \textit{Fixed-age models}. To fit with the majority of the literature studies (e.g. \citealp{Dopita+2006}, \citealp{Levesque+2010}, \citealp{Kashino+2016}), we assume the central source to be an SSP with a given metallicity (Z = 0.004, 0.008, 0.02, 0.04)  and a fixed age of 0.01 Myr. Moreover, \textcolor{black}{we adopt a grid of decreasing ionization parameters in order to simulate different ionization levels}. We assume \textcolor{black}{a grid of fixed-age ionization parameters log(U)$_0$ going from  $-3.6$ to $-2.5$ with steps of 0.1 dex}, to be consistent with the observations of unresolved star-forming H II regions (log(U)$\lesssim -2.3$, see \citealp{Yeh&Matzner2012}), local H II regions ($-3.2 < $log(U)$ < -2.9$, see \citealp{Dopita+2000}) and star-forming galaxies (see \citealp{Moustakas+2006}; \citealp{Moustakas+2010}). \textcolor{black}{In these models, given a metallicity, the shape of the ionizing source is fixed, regardless of the ionization parameter, and models with the lowest log(U)$_{0}$ can describe star-forming regions, but with very low densities of ionizing photons.}\\

\item \textit{Evolving-age models}. \textcolor{black}{We also construct models which take account of the shape variation of the ionizing SED as a function of time after SF is stopped. This evolution is illustrated in Fig. \ref{fig:Stb99_gen94} for an SSP of solar metallicity. The UV ionizing flux decreases as a function of time, with harder energies disappearing first, due to the sudden disappearence of the most massive O stars able to produce them. This behaviour is also visible in Fig. \ref{fig:nphot}}, which shows the time evolution of the number of ionizing photons below different energy thresholds (i.e. the ones relative to the emission lines studied in this work: H and O$^{+}$, 13.6 eV; O$^{++}$, 35 eV and Ne$^{++}$, 41 eV) and for different metallicities. The number of ionizing photons decreases with time, and its decline is more pronounced for harder energies and higher metallicities. Moreover, the effect of the increased metallicity is more visible at the harder energies, which are more absorbed \textcolor{black}{due to} the larger stellar opacities. \textcolor{black}{To account for the softening of the UV ionizing spectrum as a function of time after the SF quenching, in these models we simulate the central ionizing source using an SSP with given metallicity (Z = 0.004, 0.008, 0.02, 0.04) and age going from 0.01 to 10 Myr}. In particular, the youngest SSP of 0.01 Myr is taken as representative of a still star-forming region (this kind of assumption is often used in literature, e.g. \citealp{Kewley+2001}, \citealp{Dopita+2006}, \citealp{Levesque+2010}), while older SSPs are used to describe the epochs subsequent to the SF quenching. The still star-forming region can have \textcolor{black}{a ionization state described by} one of the \textcolor{black}{fixed-age ionization parameters} log(U)$_{0}$ defined before, which then evolves with time according to the $Q(H)$ time evolution (see \citealp{Rigby&Rieke2004}). Since $Q(H)$ decreases after the SF quenching, \textcolor{black}{we expect the ionization parameter to get lower as a function of time. For this reason, each model with a given age will be characterized by an evolving-age ionization parameter log(U)$_{t}$.}\\

 \end{enumerate}

For the ionized nebula, we adopt \textcolor{black}{an hydrogen density $n_{\rm{H}}=100$ cm$^{-3}$}, which is in agreement with the typical densities in observed star-forming regions (\citealp{Dopita+2000}, \citealp{Kewley+2001}, \citealp{Dopita+2006}), and the solar chemical composition by \citealp{Asplund&Grevesse2005} (the abundances of the various elements adopted in this work are listed in Table 1 of  \citealp{Dopita+2006}). \textcolor{black}{In particular, we match the metallicity of the ionized nebula with the metallicity of the ionizing stellar population\footnote{It is important to note that the solar metallicity of the Starburst99 (and BC03) models is slightly higher (Z = 0.02) than the  \citealp{Asplund&Grevesse2005} one (i.e. Z = 0.016).}}.
For non-solar metallicities, we linearly rescale the abundance of each element, except for He, C and N, for which we assume the metallicity dependences reported in \citet{Dopita+2006}. Depletion factors are fixed at the same values regardless of metallicity (e.g. \citealp{Dopita+2006}, \citealp{Nakajima&Ouchi2014}). \textcolor{black}{This implies that the dust-to-metal ratio is fixed regardless of metallicity, and that the dust-to-gas ratio is proportional to metallicity (e.g. \citealp{Issa+1990}; \citealp{Lisenfeld&Ferrara1998}; \citealp{Draine+2007}; \citealp{daCunha+2010})}. To account for the presence of dust, we adopt the default ISM grain distribution implemented in CLOUDY. However, throughout this work we always use the intrinsic fluxes provided by the CLOUDY code for all the analysed emission lines, i.e. those that not require any correction for the dust extinction.\\
Finally, since the emission lines we are interested in require very high gas kinetic temperature (i.e. $T> 20,000$ K) to be produced, we stop the calculation at the point in which the kinetic temperature of the gas has fallen down to $T\sim 4,000$ K, since at this temperature not even the hydrogen can be ionized.\\

       \section{Testing the reliability of the photoionization model}
       Before describing the behaviour of the proposed quenching diagnostics, in this section we verify the reliability of our models comparing them with data and literature. 
          
\subsection{Comparison with data}
\label{sec:valid}
   
\begin{figure}%[t!]
    \resizebox{\hsize}{!}{\includegraphics{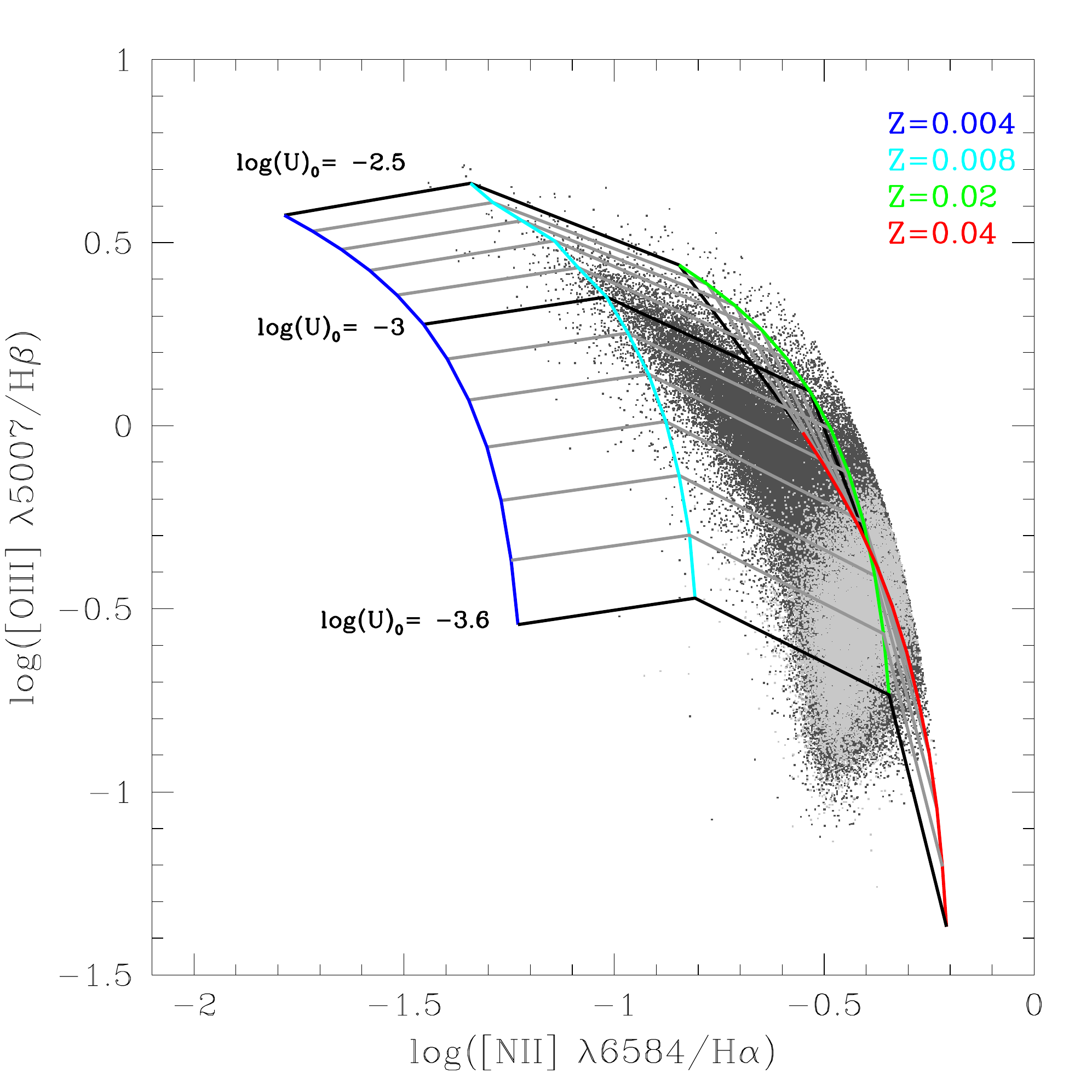}}
     \caption{Comparison between our models and observations. Dark grey points are galaxies extracted from the SDSS DR8 with S/N(H$\alpha$) > 5, S/N(H$\beta$) > 3 and S/N([N II]), S/N([O III]) > 2, while light grey points are galaxies with S/N([O III])< 2. The superimposed grid is our set of fixed-age models with different metallicities (Z = 0.004 blue; Z = 0.008 cyan; Z = 0.02 green; Z = 0.04 red) and different log(U)$_{0}$ (going from $-3.6$ to $-2.5$ with steps of 0.1 dex from bottom to top). Black curves mark the levels log(U)$_0=-3.6, -3, -2.5$, from bottom to top, as indicated.}
    \label{fig:plot_griglia_definitiva1}
    \end{figure}

    We perform the comparison with data using the Baldwin, Phillips $\&$ Terlevich diagram (BPT; \citealp{BPT1981}), which is generally adopted to distinguish star-forming from AGN ionization sources (\citealp{Veilleux&Osterbrock1987}, \citealp{Kewley+2001}, \citealp{Kauffmann+2003}, \citealp{Stasinska+2006}). To verify the consistency of our models with real data, we use a sample of $\sim 174,000$ star-forming galaxies extracted from the Sloan Digital Sky Survey Data Release 8 (SDSS DR8, e.g. \citealp{Eisenstein+2011}, see Quai et al. 2017, in prep., for details), \textcolor{black}{classified as star-forming on the basis of the BPT diagram itself, using the definition by \citet{Kauffmann+2003}}. Galaxies in this sample have $0.04\lesssim z \lesssim 0.21$, $9 \lesssim$ log(M/M$_{\sun}$) $\lesssim 12$, signal-to-noise ratio (S/N) of the H$\alpha$ flux > 5 and of the H$\beta$ flux > 3. \textcolor{black}{The threshold of S/N (H$\alpha$) has been adopted because we expect just quenched galaxy spectra to be still characterized by relatively strong low-ionization lines (see Sect. \ref{sect:approach}).}\\
    The emission line measurements for each object are derived from the MPA-JHU group \citep{Brinchmann+2004} and are corrected for dust extinction assuming the Calzetti extinction curve (\citealp{Calzetti+2000}) and using the H$\alpha$/H$\beta$ ratio to estimate the nebular colour excess $\rm{E(B-V)}$.
    \textcolor{black}{Starting from the selected sample, in the following sections we consider different subsamples of objects, selected according to the signal-to-noise ratio (S/N) of the emission lines under analysis.} Throughout the work, emission lines characterized by a S/N < 2 are regarded as not-detected lines, thus their fluxes are treated as upper limits and set to 2$\sigma$. We consider, in particular, upper limits for objects with [O III] $\lambda$5007 ([O III], hereafter) and [Ne III] undetected. On the contrary, S/N > 2 are associated to emission lines like [O II], [N II] $\lambda$6584 ([N II], hereafter) and [S II] $\lambda\lambda$6717,6731 ([S II], hereafter).\\
As illustrated in Fig. \ref{fig:plot_griglia_definitiva1}, models are able to reproduce the bulk of the data distribution in the BPT star-forming branch, with galaxies spanning the entire range of ionization parameters and metallicities considered in our study. \textcolor{black}{The discrepancy of the uppermost envelope of the SF region could be also due to residual contamination by AGN or composite sources (e.g. \citealp{Stasinska+2006}) or to shortcomings related to models. For instance, recent works have argued that, at the present state, many of the synthetic spectra generally used in the literature can produce too soft UV ionizing fluxes, especially at high metallicities (e.g. \citealp{Levesque+2010}). Moreover, first evidences that local galaxies can have higher N/O ratios and/or higher dust-to-metal ratios especially at high metallicities have been found in the very last years (\citealp{Brinchmann+2013}, \citealp{Perez-Montero+2013}, \citealp{Wu+2014}, \citealp{Morisset+2016}), suggesting that they could help in filling the gaps between models and data.
At Z > 0.02, models with different metallicities and ionization parameters overlap between each other. This can be mainly attributed to the fact that the [N II] $\lambda$6584/H$\alpha$ ([N II]/H$\alpha$, hereafter) ratio has not a smooth increase with Z, flattening at supersolar metallicities (see Sect. \ref{sec:deg} for further details). 
 \textcolor{black}{However, defining more sophisticated models able to explain the BPT diagram is beyond the aims of this work and would not influence our studies on the galaxy quenching phases.}}\\
\textcolor{black}{It is important to note that the adopted SDSS data are sensitive enough for our methodology, which basically relies on the detection of variations in emission line ratios. Our assumptions of S/N (H$\alpha$) > 5 and S/N ([O III]) $\leq$ 2 allow to detect differences between these two lines (and thus variations in their ratio) by up to a factor $\sim$ 10. More precisely, the limiting fluxes for [O III] and H$\alpha$ in our sample are $\sim0.32\times10^{-16}$ erg s$^{-1}$ cm$^{-2}$ and $\sim1.8\times10^{-16}$ erg s$^{-1}$ cm$^{-2}$, respectively, and the lowest detected log([O III]/H$\alpha$) is $\sim$ --1.5, which corresponds to a difference by a factor $\sim$ 30 between the two lines (cfr. Quai et al, 2017, in prep). Furthermore, the analyzed SDSS individual spectra have a continuum S/N around the [O III] line > 5, which has been proved to be sufficient to provide reasonable measurements of emission line fluxes (e.g. \citealp{Thomas+2013}). This further implies a good sensitivity to variations in ratios involving emission lines in this spectral region.}
%the high S/N of the H$\alpha$ line ensures the weakness of the high ionization lines of individual SDSS spectra to be real and not related to selection effects (e.g. if H$\alpha$ was weak).
 
 \subsection{Comparison with other models}
\begin{figure}%[t!]
    \resizebox{\hsize}{!}{\includegraphics{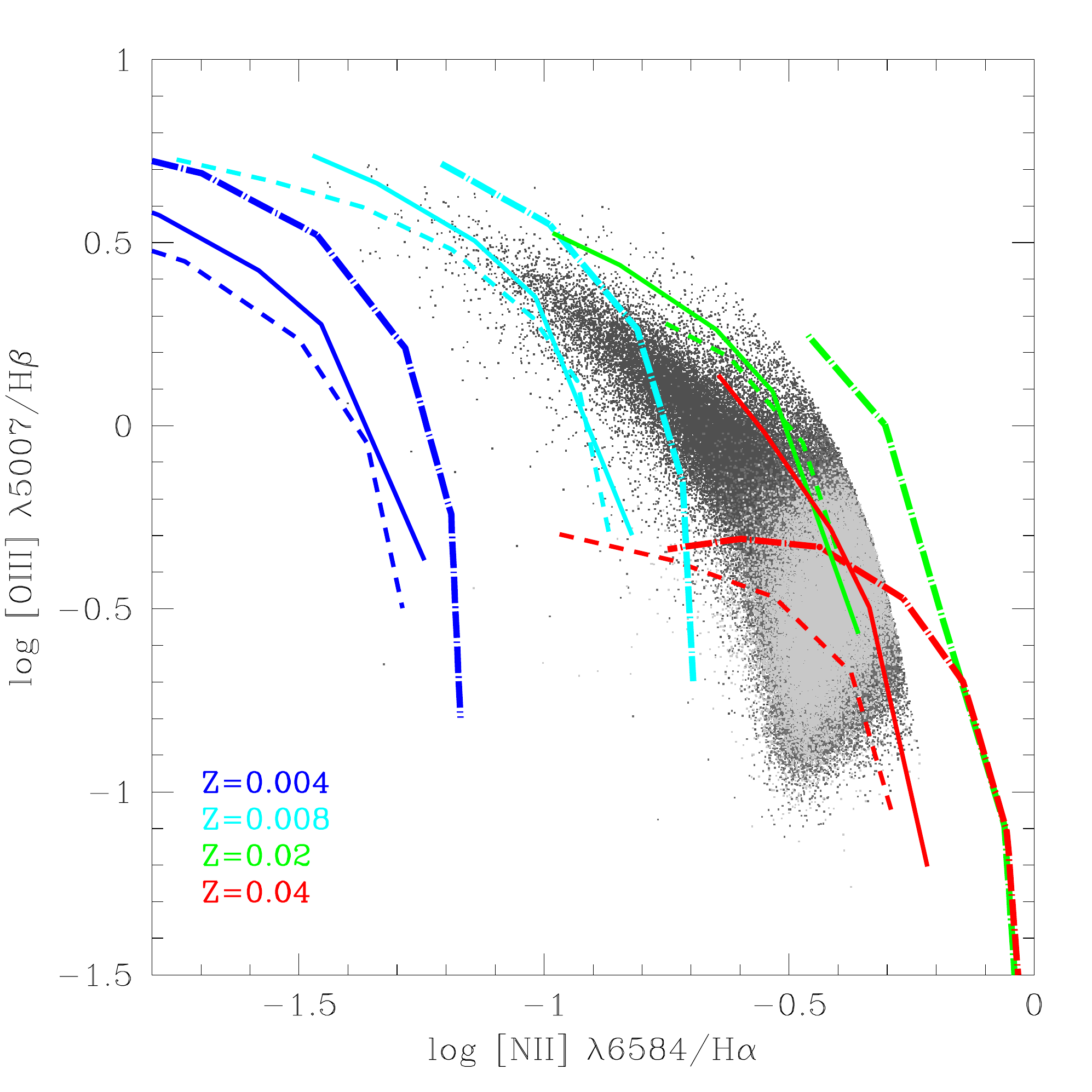}}
      \caption{Comparison among our models (solid curves), \citet{Levesque+2010} (dashed curves) and \citet{Kewley+2001} (dotted-dashed curves) predictions, for $-3 < $log(U)$_{0} < -2$. Grey points are the sample extracted from the SDSS DR8, colour coded as in Fig. \ref{fig:plot_griglia_definitiva1}. Different colours indicate different metallicities (Z = 0.004, blue; Z = 0.008 cyan; Z = 0.02, green; Z = 0.04, red).}
    \label{fig:plot_confronti}
    \end{figure}

\textcolor{black}{We verify the reliability of our models comparing them with other predictions available in the literature: i.e. those presented in \citet{Kewley+2001}  (Kew01, hereafter) and those by \citet{Levesque+2010} (Lev10, hereafter). In Table 1 we summarize the main parameters characterising Kew01, Lev10, and our model.}
%We verify the reliability of our models also comparing them with other model predictions available in the literature. In particular, we consider \citet{Kewley+2001} (Kew01, hereafter) and \citet{Levesque+2010} (Lev10, hereafter) model predictions 
% which adopt SSPs as central ionizing sources of the star-forming regions and assume sets of ingredients similar to the ones adopted in this work 
 %\textcolor{black}{(Table \ref{tab:ingr} summarizes the main ingredients adopted in this work and in the literature).} % Briefly, \citet{Levesque+2010} models are obtained with the MAPPINGS III photoionization code (\citealp{Sutherland&Dopita1993}, \citealp{Groves+2004b}), adopting fixed-age Starburst99 SSPs of 0.01 Myr and $M=10^6 M_{\sun}$, Geneva-HIGH 1994 evolutionary tracks, Paldruach/Hillier stellar atmospheres (\citealp{Pauldrach+2001}; \citealp{Hillier&Miller1998}), ionization parameters going from log(U)$_{0}= -3.5$ to log(U)$_{0}= -1.9$, \textcolor{black}{and metallicities Z = 0.001, 0.004, 0.008, 0.02, 0.04}. The \citet{Kewley+2001} models are produced with MAPPINGS III, fixed-age (i.e. 0.01 Myr) Starburst99 SSPs, Geneva-HIGH 1994 tracks, Lejeune -- Schmutz stellar atmospheres, ionization parameters  $-3.7< $log(U)$_{0} < -2$. 
Fig. \ref{fig:plot_confronti} shows the results within the BPT plane. 
% for $-3 < $log(U)$_{0}< -2$ and metallicities Z = 0.004, 0.008, 0.02, 0.04.
Note that, to be consistent with the literature, this comparison is performed only in the case of fixed-age models. \\
Our models are in reasonable agreement with the literature ones, at each metallicity. However, at Z $\geq0.02$, the dispersion among different predictions is more pronounced, and in some cases (e.g. Kew01) models produce too high [N II]/H$\alpha$ values with respect to the data. 
\textcolor{black}{We also find that, at Z = 0.04, a difference in the [O III]/H$\beta$ vs. [N II]/H$\alpha$ slope is present. This can be due to the different dust-prescriptions adopted, which lead our models to be dustier than those of Kew01 and Lev10. The discrepancy is larger at higher metallicities, where our models have an higher dust-to-gas ratios with respect to those of the literature (see Sect. \ref{sec:photoionmodel}). 
%The discrepancy is expected to affect higher metallicities the most, where our models are characterized by larger dust-to-gas ratios and thus by larger dust amounts (see Sect. \ref{sec:photoionmodel}).
In particular, when many ionizing photons are provided by the central source (i.e. at high log(U)$_0\sim-2$), the larger amount of dust in our models amplifies grain photoelectric heating processes, which produce free electrons able to collisionally- excite the [O III]. This increases the [O III]/H$\beta$ ratio more than the Kev01 and Lev01 models, hence producing a steeper slope for our models. At lower log(U)$_0$, when grain photoelectric heating processes decrease because of the very low number of ionizing photons, there is instead a better agreement among different models.\\
%In particular, the slope difference is generated by the fact that, when many ionizing photons are provided by the central source (i.e. at high log(U)$_0\sim-2$), the larger amount of dust in our models amplifies grain photoelectric heating processes, which produce free electrons able to collisionally-excite the [O III] increasing the [O III]/H$\beta$ ratio more than the Kev01 and Lev01 models. At lower log(U)$_0$, when grain photoelectric heating processes \textcolor{black}{decrease} because of the very low number of ionizing photons, there is instead a better agreement among different models. \\ 
Note that this does not affect the lowest log(U)$_0$ regimes (i.e. log(U)$_0\lesssim-3.5$), which are the ones relevant in our study.}
%Note that this discrepancy does not affect the lowest log(U)$_{0}$, which are the ones we are interested in.
%We do not expect this discrepancy to affect the predicted emission line ratios time evolution, which is basically independent from their value at the time at which the quenching begins (see Sect. \ref{sec:que_diagn} for further details).
\textcolor{black}{More generally, the discrepancies among different models are related to the different ingredients and photoionization codes adopted for the simulations. In particular, as already mentioned in Sect. \ref{sec:valid}, different stellar atmospheres can produce UV SEDs with different slopes, affecting the emission line ratios which are more sensitive to the UV hardness (\citealp{Schmutz+1992}, \citealp{Hillier&Miller1998}, \citealp{Dopita+2000}, \citealp{Pauldrach+2001}, \citealp{Kewley+2001}, \citealp{Levesque+2010}). Furthermore, Kew01 models adopt higher depletion factors for some elements (e.g. C and Fe) and thus higher dust-to-metal ratios for a given metallicity. This produces an increase of the gas electron temperature, which favours the cooling from metal optical lines like [O III] and [N II] (e.g. \citealp{Shields&Kennicutt1995}, \citealp{Charlot&Longhetti2001}; \citealp{Brinchmann+2013}), increasing their fluxes and thus the emission line ratios in which they are involved.}

% \begin{center}
%\begin{table*}[h]
%{\small
%\hfill{}
%\caption{Ingredients of \citet{Kewley+2001} and \citet{Levesque+2010} models.}
%\renewcommand{\arraystretch}{1.4}
%\begin{tabular}{lccc} 
 %\hline
%&Kew01 & Lev10\\
%\hline
%Code& MAPPINGS III& MAPPINGS III\\
 %Ev. tracks&Geneva High 1994& Geneva High 1994\\
 %Stellar atm&Lejeune-Shmutz& Paldruach-Hillier\\
 % log(U)$_{0}$&$-3.7 <$log(U)$_{0}< -2$&$-3.5 <$log(U)$_{0}< -1.9$ \\
   %Age &0.01 Myr& 0.01 Myr \\
   %Z&0.001, 0.004, 0.008, 0.02, 0.04&0.004, 0.008, 0.02, 0.04 \\
%\hline

%\label{tab:ingr}
%\end{tabular}}
%\end{table*}
%\end{center}
   
   \begin{center}
\begin{table*}[h]
{\small
\hfill{}
\caption{Main ingredients of our models, \citet{Kewley+2001}'s and \citet{Levesque+2010}'s ones.}
\renewcommand{\arraystretch}{1.4}
\begin{tabular}{lcccc} 
 \hline
&This work&Kew01 & Lev10\\
\hline
Code& CLOUDY 13.03&MAPPINGS III& MAPPINGS III\\
 Evolutionary tracks&Geneva High 1994&Geneva High 1994& Geneva High 1994\\
 Stellar atmospheres&Lejeune-Shmutz&Lejeune-Shmutz& Paldruach-Hillier\\
  log(U)$_{0}$&$-3.6 <$log(U)$_{0}< -2$&$-3.7 <$log(U)$_{0}< -2$&$-3.5 <$log(U)$_{0}< -1.9$ \\
   Ionizing source SFH & SSP &SSP&SSP\\
   Ionizing source age &0.01 Myr&0.01 Myr& 0.01 Myr \\
   Metallicity &0.001, 0.004, 0.008, 0.02, 0.04&0.004, 0.008, 0.02, 0.04&0.004, 0.008, 0.02, 0.04 \\
\hline

\label{tab:ingr}
\end{tabular}}
\end{table*}
\end{center}

 \section{Quenching diagnostics}

Before describing how the two emission line ratios mentioned above can help in identifying galaxies which are quenching their SF,  in the following sections we describe in more details their behaviour as a function of time, ionization parameter and metallicity, and the relative degeneracies. \\

\subsection{Emission lines ratios and their evolution with time}
\label{sec:que_diagn}
\begin{figure}%[t!]
       \resizebox{\hsize}{!}{\includegraphics{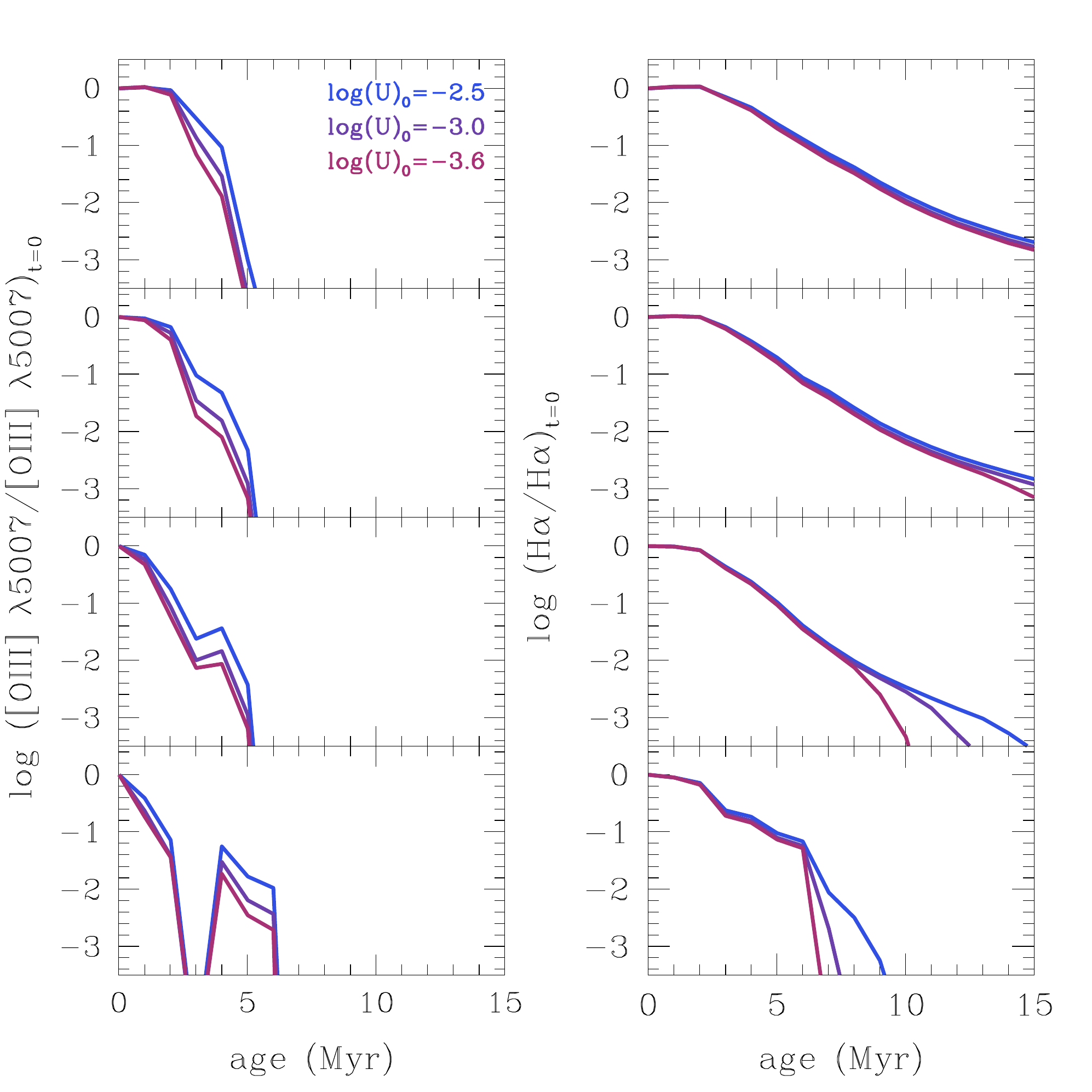}}
   \caption{Evolution of the line luminosity relative to the initial one at $t=0$ for [O III] (left) and H$\alpha$ (right) as a function of time, metallicity and log(U)$_{0}$. Metallicity (Z = 0.004, 0.008, 0.02, 0.04) increases from the top to the bottom panel. In each panel, we show the results for log(U)$_{0}$ --2.5, --3, --3.6, with log(U)$_{0}$ decreasing from blue to red, as indicated.}
   \label{fig:oiii_Ha_age_met}
     \end{figure}

\begin{figure}%[t!]
     \resizebox{\hsize}{!}{\includegraphics{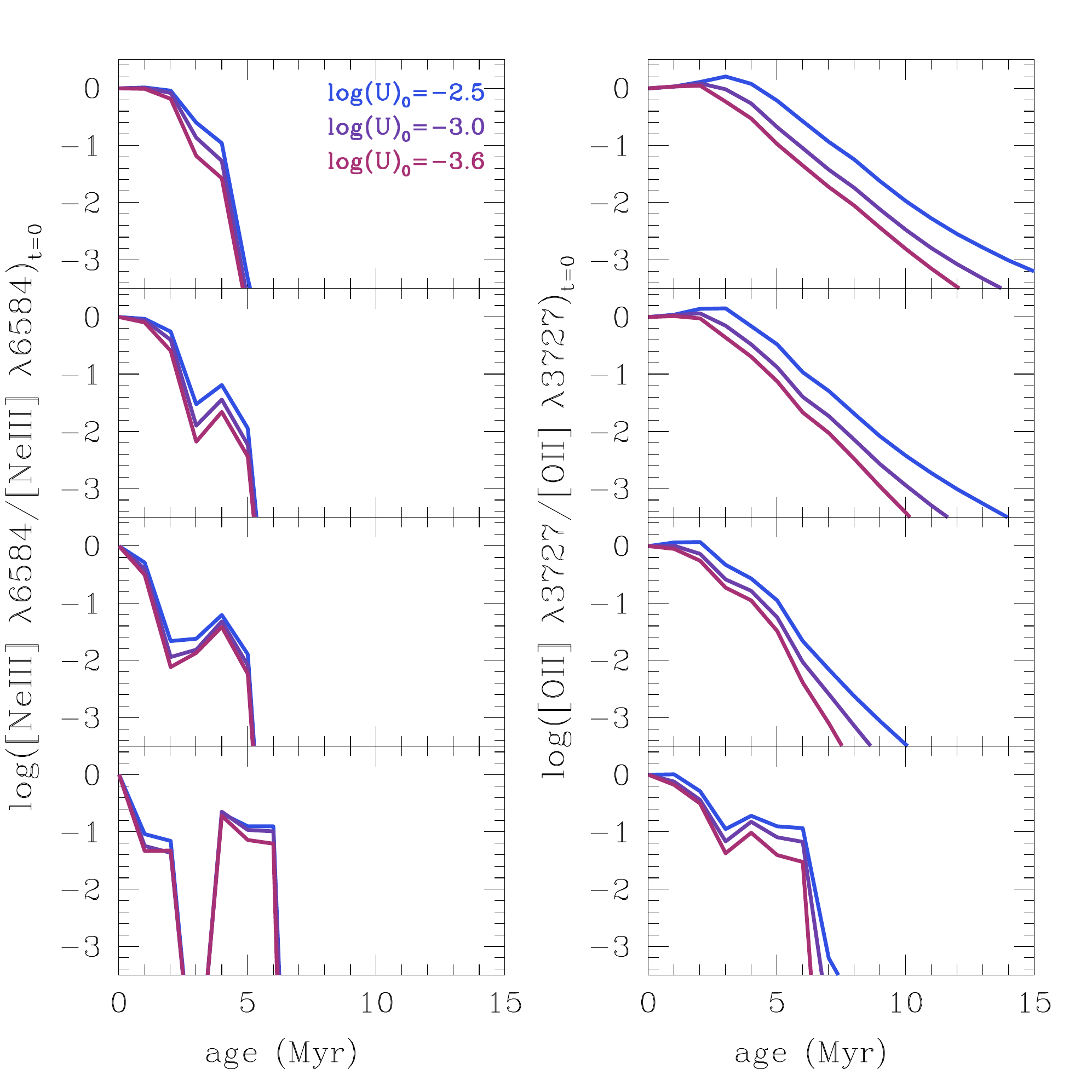}}
     \caption{Percentage evolution of [Ne III] (left) and [O II] (right) as a function of time, metallicity and log(U)$_{0}$. Metallicity (Z = 0.004, 0.008, 0.02, 0.04) increases from the top to the bottom panel. In each panel, we show the results for log(U)$_{0}$ --2.5, --3, --3.6, with log(U)$_{0}$ decreasing from blue to red, as indicated.}
     \label{fig:neiii_oii_age_met}
    \end{figure}

The two proposed emission line ratios are expected to suddenly react to the halt of the SF within a star-forming region. In this regard, Figs. \ref{fig:oiii_Ha_age_met} and \ref{fig:neiii_oii_age_met} show the percentage variation of the emission lines as a function of time from the SF quenching, \textcolor{black}{for different metallicities and initial log(U)$_{0}$}. Regardless of the initial log(U)$_{0}$, high-ionization lines behave differently from low-ionization ones, with the former changing more significantly as a function of the various parameters. For instance, [O III] and [Ne III] drop by more than a factor 10 within the first $2-3$ Myr after the SF quenching, especially at the highest metallicity and the lowest log(U)$_{0}$. Moreover, they experience a temporary rise at $\sim$ 4 Myr, which is particularly pronounced at the highest metallicities. This rise can be attributed to Wolf-Rayet (WR) stars, which are more numerous at higher metallicities (\citealp{Schaller+1992}; see also \citealp{Levesque+2010}) and can supply, on short time intervals, very energetic photons able to re-ionize O$^{++}$ and Ne$^{++}$, producing a temporary increase of the line fluxes. \textcolor{black}{After $5-6$ Myr from the SF quenching, for all metallicities, both [O III] and [Ne III] have declined by a factor $\sim$ 100 with respect to their values in the star-forming phase and, at $\sim$ 10 Myr they are more than a factor $\sim$ 1000 lower than their initial value.\\
\textcolor{black}{Compared to [O III] and [Ne III], H$\alpha$ and [O II] have a slower decline as a function of time, which is also delayed with respect to high-ionization lines. [O II], which differently from H$\alpha$ depends on metallicity, is also less dependent on it than [O III] and [Ne III].} At $5-6$ Myr, when the high-ionization lines definitely drop, [O II] and H$\alpha$ have declined by a factor $\sim$ 10 less, regardless of metallicity. In particular, it takes $\sim$ $10-15$ Myr for the low-ionization lines to decline by a factor $\sim 100-1000$. \textcolor{black}{These predictions confirm that just quenched galaxies could lack of high-ionization emission lines, still having relatively strong low-ionization lines in their spectra.}\\ 
}

\begin{figure}%[t!]
     \resizebox{\hsize}{!}{\includegraphics{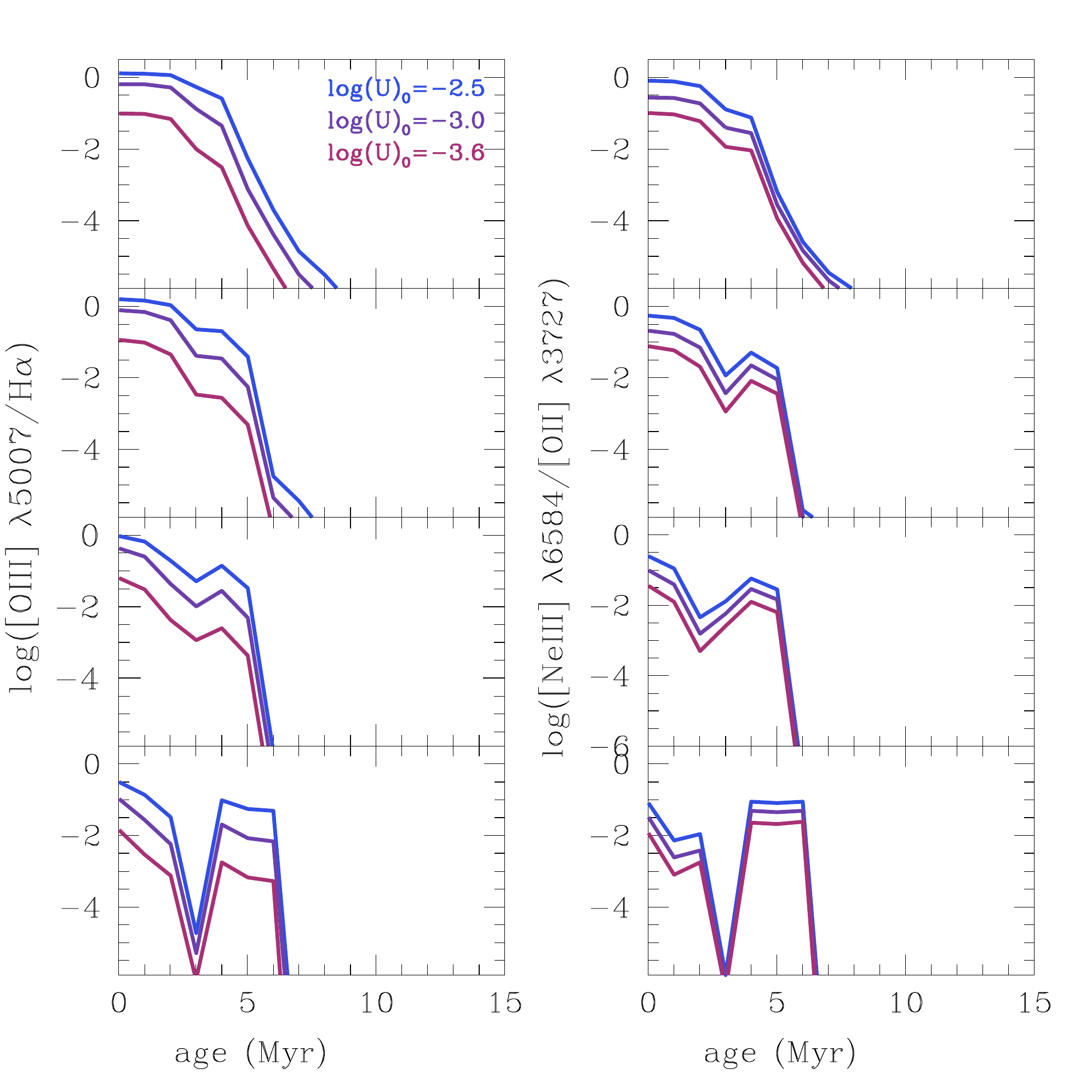}}
         \caption{[O III]/H$\alpha$ (left) and [Ne III]/[O II] (right) evolution as a function of time, metallicity and log(U)$_{0}$. Metallicity (Z = 0.004, 0.008, 0.02, 0.04) increases from the top to the bottom panel.  In each panel, we show the results for log(U)$_{0}$ --2.5, --3, --3.6, with log(U)$_{0}$ decreasing from blue to red, as indicated.}
          \label{fig:oiii_Ha_neiii_oii_age_met}
    \end{figure}

\begin{figure}%[t!]
     \resizebox{\hsize}{!}{\includegraphics{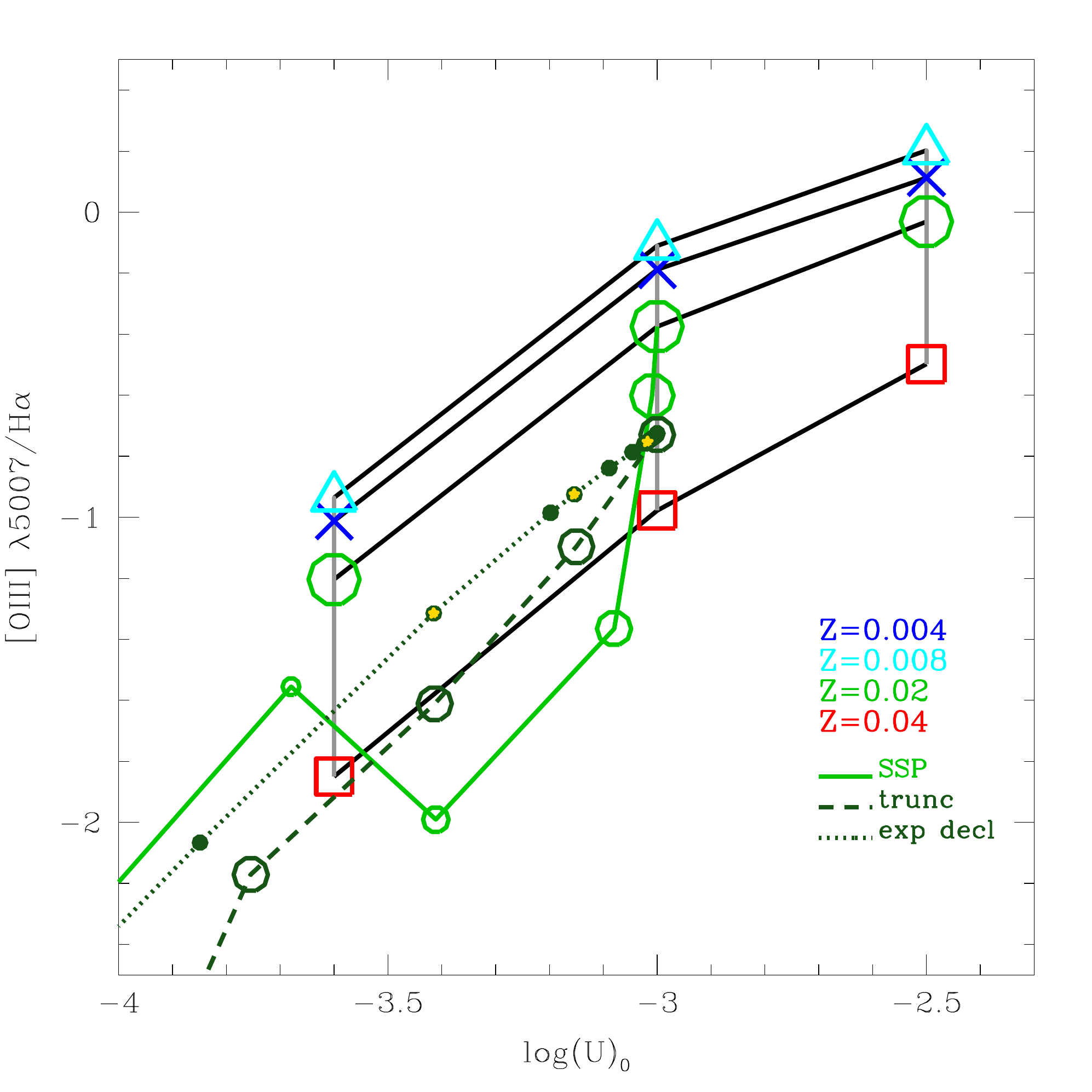}}
       \caption{\textcolor{black}{[O III]/H$\alpha$ as a function of log(U)$_0$ and log(U)$_t$. Grey curves connect models with the same log(U)$_0$, for the three log(U)$_0= -3.6, -3, -2.5$ and different metallicity (Z = 0.004 blue; Z = 0.008 cyan; Z = 0.02 green; Z = 0.04 red), while black curves connect models with the same metallicity. %Colored curves associated to different symbols are SSP evolving-age models with an initial log(U)$_0= -3$, for the four considered metallicities and with a time step of 1 Myr. 
 For Z = 0.02, evolving-age models for SSP (light green empty circles), truncated (dark green empty circles) and the exponentially declining (dark green filled circles) SFHs are shown (see Sect. \ref{sec:sfh} for further details), for an initial log(U)$_0= -3$. The emission line ratio evolution is illustrated with a time step of 1 Myr within the first 10 Myr after quenching, $\sim$ 20 Myr from 10 to 100 Myr after quenching, and 100 Myr even further. For the exponentially declining SFH, \textcolor{black}{gold small stars mark the values of the emission line ratios corresponding to 10, 80, and 200 Myr after the SF quenching, from the highest to the lowest value of [O III]/H$\alpha$.}}}
     \label{fig:plot_oiii_ha_logU}
    \end{figure}

  The behaviour of individual emission lines is reflected in the two emission line ratios considered in our study, as illustrated in Fig. \ref{fig:oiii_Ha_neiii_oii_age_met}. In particular, both [O III]/H$\alpha$ and [Ne III]/[O II] decrease for increasing age and metallicity, at each log(U)$_{0}$. For both emission line ratios, the decline is more pronounced at higher metallicities, \textcolor{black}{since for a given age, more metallic massive stars have softer UV spectra}. Indeed, at log(U)$_{0}=-3$ and at the lowest Z, [O III]/H$\alpha$ and [Ne III]/[O II] decrease by $\sim$ 0.1 dex and $\sim$ 0.2 dex within $\sim$ 2 Myr from the SF quenching respectively, while at the highest Z they drop by $\sim$ 1.3 dex and $\sim$ 1 dex, within the same time interval. However, regardless of metallicity, the two emission line ratios are characterized by a decline by a factor $\sim$ 10 within $\sim 10$ Myr from the epoch of the SF quenching. \textcolor{black}{Although fixed-age models assume a fixed shape for the SED of the ionizing source, the ones which are characterized by the lowest log(U)$_0$ can in some sense describe star-forming regions that are quenching their SF, since low-ionization parameters are related to low numbers of ionizing photons and then to low levels of star formation. For this reason, it can be interesting to investigate the behaviour of the emission line ratios under analysis as a function of both log(U)$_{0}$ and log(U)$_{t}$. Fig. \ref{fig:plot_oiii_ha_logU} illustrates the case of [O III]/H$\alpha$. \textcolor{black}{For Z=0.02, a decline by a factor $\sim$ 10 of this ratio corresponds to a decrease in log(U)$_{t}$ by 0.1 dex (starting from an initial log(U)$_0=-3$), within a time interval of $\sim$ 2 Myr, while only a more pronounced decrease by $\sim$ 1 dex in log(U)$_{0}$ can produce the same effect. Therefore, the decline of [O III]/H$\alpha$ is more rapid for evolving-age models than for fixed-age ones. This is due to the fact that the former include the additional effect of the UV softening as a function of time.}\\
 % A decline by a factor $\sim$ 10 of this ratio corresponds to a decrease in log(U)$_{t}$ by 0.1 -- 0.4 dex (starting from an initial log(U)$_0=-3$), within a time interval of 2 -- 10 Myr, depending on metallicity, while only a more pronounced decrease by $\sim$ 1 dex in log(U)$_{0}$ can produce the same effect. Therefore, the decline of [O III]/H$\alpha$ is more rapid for evolving-age models than for fixed-age ones. This is due to the fact that the former include the additional effect of the UV softening as a function of time.\\
}

%As already mentioned in Sect. \ref{sec:photoionmodel}, for all models we assume an hydrogen density $n_{\rm{H}}=$ 100 cm$^{-3}$ for the ionized nebula.\\
\subsection{\textcolor{black}{The influence of different hydrogen densities}}

\begin{figure}%[t!]
     \resizebox{\hsize}{!}{\includegraphics{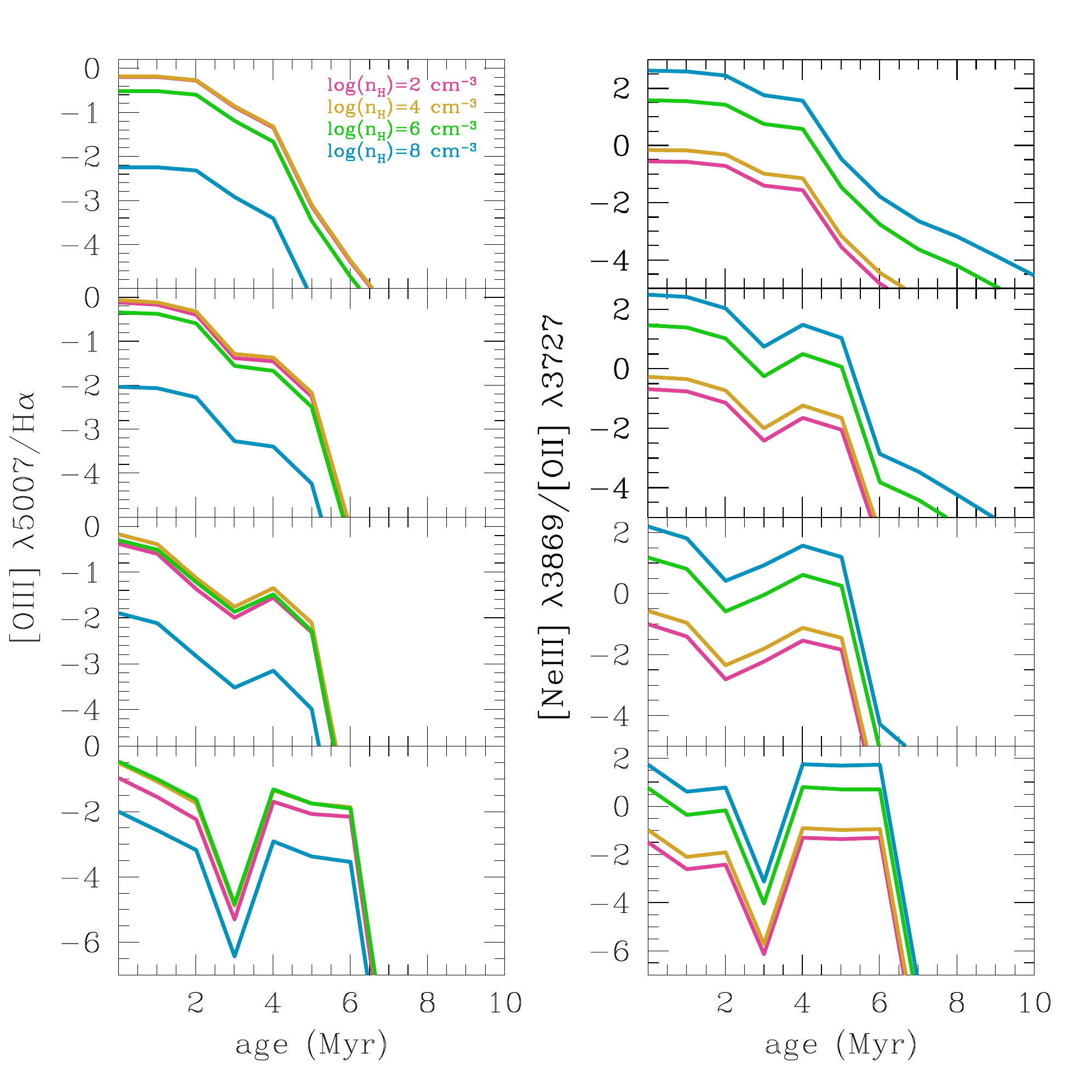}}
       \caption{\textcolor{black}{[O III]/H$\alpha$ and [Ne III]/[O II] time evolution for different values of n$_{\rm{H}}$, and log(U)$_{0}$= -- 3. The hydrogen density increases from log(n$_{\rm{H}}$)=2 (which is our default value) to log(n$_{\rm{H}}$)=8, from pink to cyan, as labelled. Metallicity (Z = 0.004, 0.008, 0.02, 0.04) increases from the top to the bottom panel.}}
     \label{fig:den}
    \end{figure}

\textcolor{black}{
The typical hydrogen densities of H II regions are of the order of $n_{\rm{H}}$$~\sim$ 100 cm$^{-3}$, and are therefore well below the critical densities of the forbidden lines considered in our analysis\footnote{(${\rm{n_{crit}}}_{\rm{[O III]}} \sim 7 \times10^{5}$ cm$^{-3}$, $\rm{{n_{crit}}}_{\rm{[Ne III]}} \sim 1.1\times10^7$ cm$^{-3}$), $\rm{{n_{crit}}}_{\rm{[O II]}} \sim 1\times10^4$ cm$^{-3}$).}. However, it can be interesting to investigate how [O III]/H$\alpha$ and [Ne III]/[O II] are affected by an increase of n$_{\rm{H}}$ towards the critical densities (${\rm{n_{crit}}}$) of [O III], [Ne III] and [O II]. \\ As illustrated in Fig. \ref{fig:den}, we find that [O III]/H$\alpha$ increases by only $\sim$ 0.2 dex for $n_{\rm{H}}$ < $\rm{n_{crit}}$([O III]), i.e. when H$\alpha$ and [O III] are both increasing due to the increment of ionized hydrogen and free electrons able excite metal ions. On the contrary, when $n_{\rm{H}}$ > $\rm{n_{crit}}$([O III]), [O III]/H$\alpha$ rapidly drops, due to the prevalence of collisional on radiative de-excitations for the [O III] line.\\
The [Ne III]/[O II] ratio has a more complex behaviour, since it involves two forbidden lines. In particular, it increases by $\sim$ 0.5 dex for $n_{\rm{H}}$ < $\rm{n_{crit}}$([O II]), while it starts to rapidly increase when $\rm{n_{crit}}$([O II]) is reached and [O II] saturates. This trend is inverted when also $\rm{n_{crit}}$([Ne III]) is reached.  \\
In general, we find that zero-age models with high-n$_{\rm{H}}$ can produce the same [O III]/H$\alpha$ values as quenching ones with low-n$_{\rm{H}}$ (and zero age models with low-n$_{\rm{H}}$ can produce the same [Ne III]/[O II] values as quenching ones with high-n$_{\rm{H}}$), at each metallicity. This means that our models can in principle be affected by the n$_{\rm{H}}$ degeneracy. However, this holds true for n$_{\rm{H}}\gtrsim4$, which is much larger than the typical n$_{\rm{H}}$ in H II regions. This allows to exclude the very low [O III]/H$\alpha$ and [Ne III]/[O II] found in this work to be related to density rather than quenching effects. Finally, we notice that the sudden drop of the emission line ratios after a few Myr from the SF quenching is present regardless of n$_{\rm{H}}$.
%This means that our models can in principle be affected by the $n_{\rm{H}}$ degeneracy, but at $n_{\rm{H}}\gg 100$ cm$^{-3}$, which excludes that the very low [O III]/H$\alpha$ and [Ne III]/[O II] values found in this work could be related to density rather than quenching effects.
%Finally, we notice that the sudden drop of the emission line ratios after a few Myr from the SF quenching is always present, regardless of n$_{\rm{H}}$.
}

%This value is well below the critical densities $\rm{n_{crit}}$ of [O III] and [Ne III]\footnote{(${\rm{n_{crit}}}_{\rm{[O III]}} \sim 7 \times10^{5}$ cm$^{-3}$, $\rm{{n_{crit}}}_{\rm{[Ne III]}} \sim 1.1\times10^7$ cm$^{-3}$)}, which excludes that, at a given metallicity, the very low values of [O III]/H$\alpha$ and [Ne III]/[O II] could be related to density effects rather than quenching. \textcolor{black}{In particular, we find that, for $n_{\rm{H}}< n_{\rm{crit}}$, an increase of $n_{\rm{H}}$ leads to an increase of the analysed emission line ratios, due to the increment of free electrons able to excite the metal ions producing forbidden lines. Conversely, at $n_{\rm{H}}\sim n_{\rm{crit}}$, forbidden lines start to decrease, due to the prevalence of collisional on radiative de-excitations. Finally, we verified that the sudden drop of the emission line ratios after a few Myr from the SF quenching is always present, regardless of $n_{\rm{H}}$.} 

  \subsection{The influence of different synthetic stellar spectra}
  
\begin{figure}%[t!]
    \resizebox{\hsize}{!}{\includegraphics{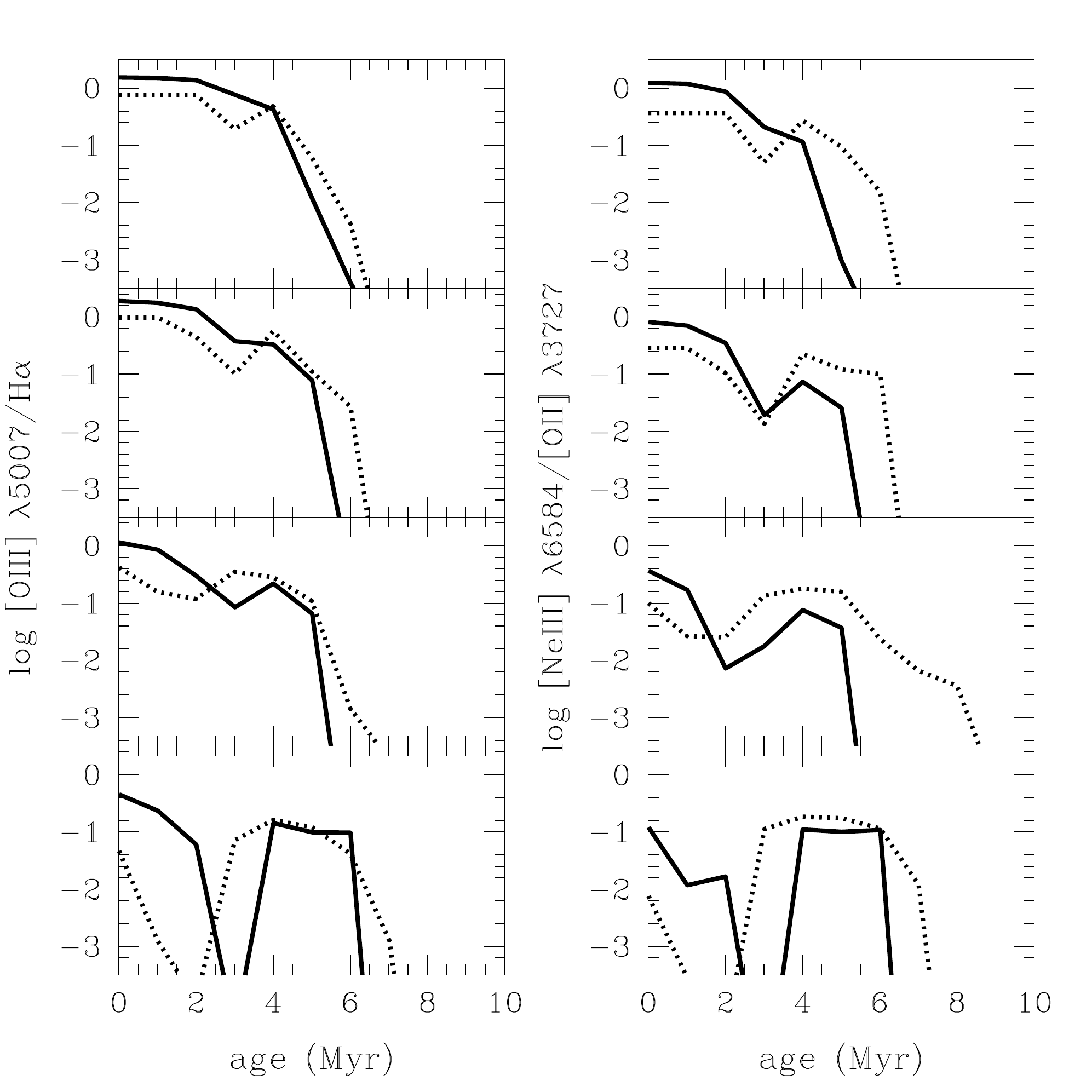}}
       \caption{Comparison between [O III]/H$\alpha$ and [Ne III]/[O II] obtained assuming Starburst99 (solid curves) and BC03 (dotted curves) models to simulate the central ionizing source, assuming log(U)$_{0}=-3$. Metallicity (Z = 0.004, 0.008, 0.02, 0.04) increases from the top to the bottom panel.}
    \label{fig:comp_stb99_bc03}
    \end{figure}

     In this section we verify if the results illustrated in Sect. \ref{sec:que_diagn} are influenced by the use of different synthetic spectra to describe the central ionizing source. Fig. \ref{fig:comp_stb99_bc03} illustrates a comparison between Starburst99 and BC03 models, in the case of log(U)$_{0}=-3$ and for the four metallicities considered in this work. The general behaviour of [O III]/H$\alpha$ and [Ne III]/[O II] obtained using BC03 models is in agreement with the ones resulting from Starburst99 spectra. \textcolor{black}{Both emission line ratios decline by a factor $\sim$ 10 within $\sim$ 10 Myr from the quenching of the SF, at each metallicity. After the SF shutdown, the behaviour of the two emission line ratios is more uneven in the case of BC03 models than for Starburst99 ones. This is probabily due to the different stellar atmospheres (see \citealp{Charlot&Longhetti2001}) and evolutionary tracks (see \citealp{Bruzual&Charlot2003}) used to compute the models. For instance, BC03 models are based on the Padova1994 evolutionary tracks \citep{Bressan+1994}, which are characterized by a higher number of WR stars with respect to the Geneva-HIGH 1994 ones (as discussed in \citealp{Bruzual&Charlot2003}). This can give more pronounced discontinuities in the emission line ratios at the very early ages after the SF quenching.} \textcolor{black}{The time at which the two emission line ratios begin to definitely drop is the same in the case of BC03 and Starburst99 spectra.  After this epoch, the time required to have a decrease by a factor $\sim$ 1000 is slightly higher for BC03 models, with a shift of $\sim1-3$ Myr for both emission line ratios. Apart from these small differences, we can conclude that our methodology is valid also when BC03 models are adopted. This is an interesting result, since BC03 models are generally used only to describe more advanced evolutionary phases of galaxy evolution (e.g. \citealp{Leitherer+1999}, \citealp{Bruzual&Charlot2003}, \citealp{Chen+2010}).}

\subsection{The influence of different star formation histories} 
\label{sec:sfh}
\begin{figure}%[t!]
    \resizebox{\hsize}{!}{\includegraphics{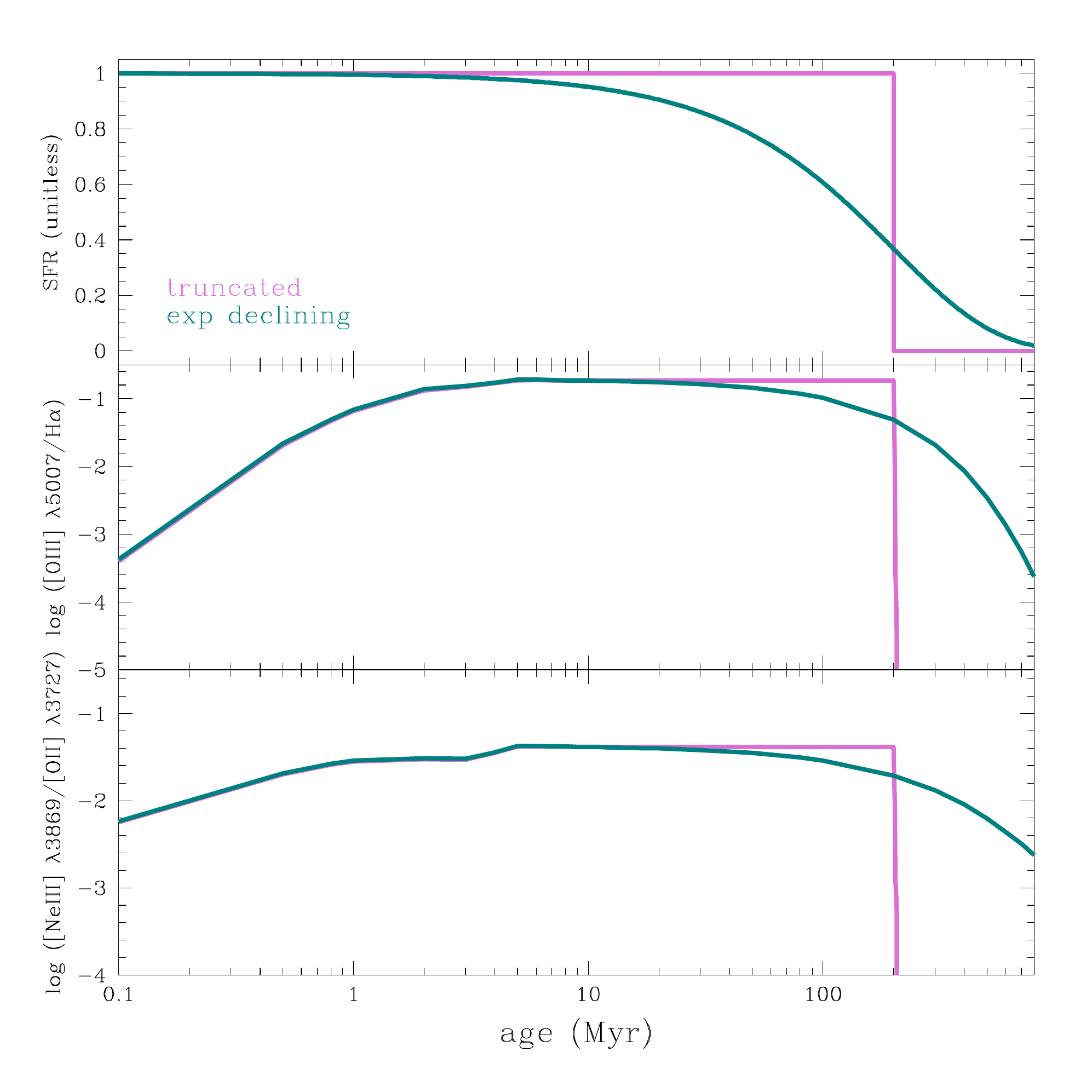}}
    \caption{SFR, [O III]/H$\alpha$ and [Ne III]/[O II] emission line ratios as a function of time for the truncated and the exponentially declining SFHs, Z = 0.02 and $t=0$. In the top panels, the SFRs have different scales due to the different definitions of the two SFHs (described in the text). In both cases of truncated and exponentially declining SFH, the fast rise before the time of quenching is visible, as described in the text.}
                 \label{fig:sfh1}
    \end{figure}

\begin{figure}%[t!]
    \resizebox{\hsize}{!}{\includegraphics{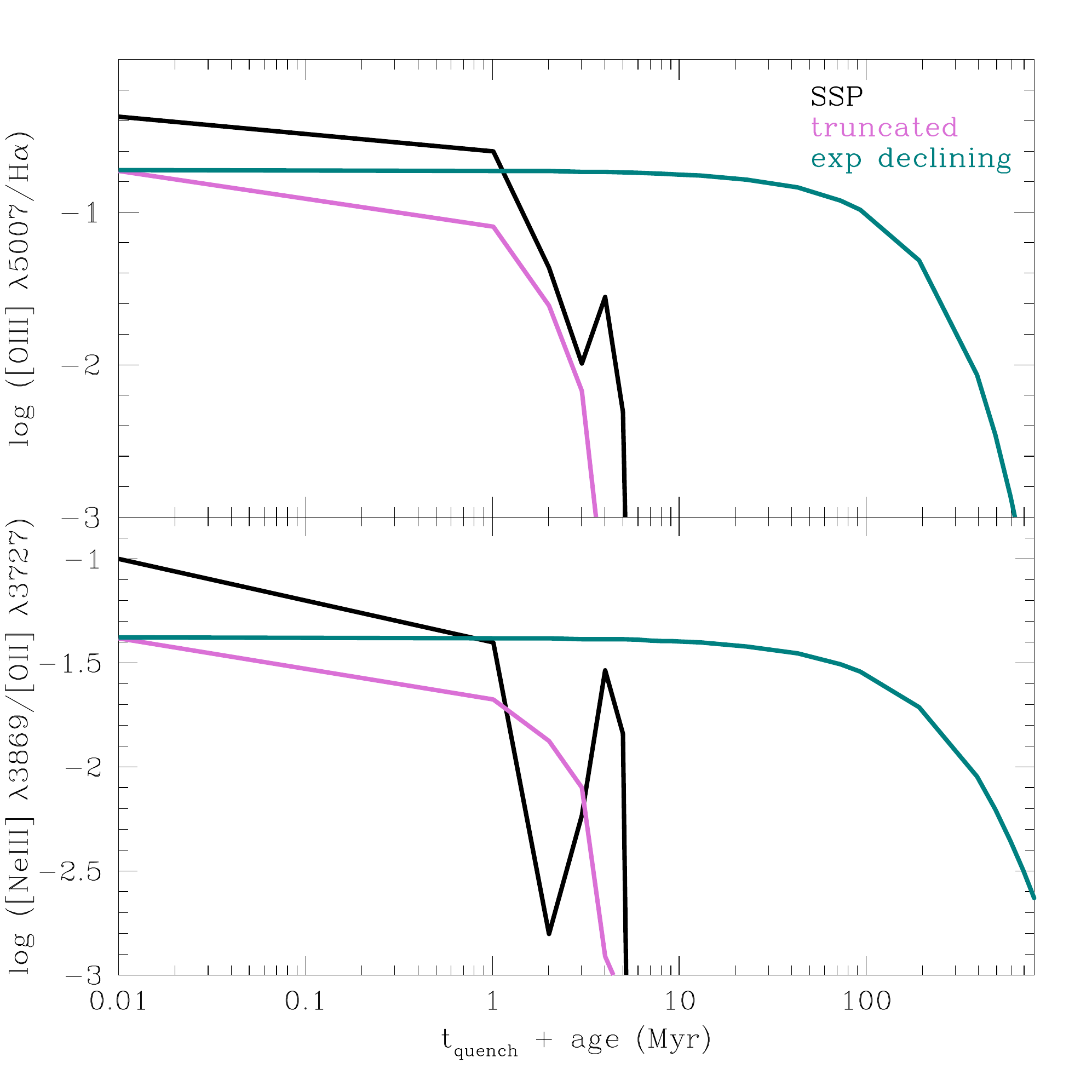}}
    \caption{\textcolor{black}{[O III]/H$\alpha$ and [Ne III]/[O II] emission line ratios as a function of time for different SFHs and Z = 0.02. The SSP (black curve), truncated (violet curve) and the exponentially declining (green curve) SFHs are shown up to $\sim800$ Myr from the time of quenching (indicated as $t_{\rm{quench}}$).}}
                 \label{fig:sfh2}
    \end{figure}

\textcolor{black}{The instantaneous SF quenching modelled so far can be considered an extreme case, which allows to better show the strength of our approach to probe SF quenching on very short time-scales.} In this section we investigate how the behaviour of [O III]/H$\alpha$ and [Ne III]/[O II] changes if different star-formation histories (SFHs) and if a \textcolor{black}{more realistic} and smoother SFHs are assumed. In particular, using Starburst99 models, we analyse the case of a truncated SFH, with SFR = 1 $M_{\sun}$ yr$^{-1}$ up to 200 Myr and zero at older ages, and the case of a smoother decline described by an exponentially declining SFR (i.e. SFR $\propto e^{-t/\tau}$ ), with $\tau=$ 200 Myr, since this is the SFH shape generally assumed to describe local star-forming galaxies (e.g. \citealp{Bell&DeJong2001}). \textcolor{black}{For the truncated SFH, a SFR = 1 $M_{\sun}$ yr$^{-1}$ is chosen to match the typical SFRs of SDSS star-forming galaxies (\citealp{Brinchmann+2004}, \citealp{Whitaker+2014}) at stellar masses comparable with the ones of our sample}. For the exponentially declining SFH, \textcolor{black}{models are normalized to $10^{6}~M_{\sun}$. Moreover,} we do not consider very high values of $\tau>200$ Myr, since in this case the SFH would extend at much larger times, incompatible with our assumption that galaxies are quenching their SF\textcolor{black}{. Furthermore, larger $\tau$'s have been demonstrated to produce galaxies which never leave the blue cloud (e.g. \citealp{Schawinski+2014}).}\\
For both SFHs, we consider Z = 0.02 and an initial ionization level log(U)$_{0}=-3$. In particular, to be consistent with the SSP case (see Sect. \ref{sec:que_diagn}), for the truncated SFH we attribute log(U)$_{0}=-3$ at the time corresponding to the last act of SF (i.e. 200 Myr in this case). For the exponentially declining SFH, log(U)$_{0}=-3$ is instead associated to the age at which $Q(H)$ begins to drop ($\sim 10$ Myr), since we are interested in the epochs at which the SF starts to quench. \textcolor{black}{Moreover, these choices allow to avoid, at very early times $t<t_{\rm{MS}}$(O stars) (i.e. $\sim2\times10^7$ yr), the short transient phase during which a rapid rise of ionizing photons characterizes both SFHs (see \citealp{Madau+1998}), \textcolor{black}{also reflecting on the two emission line ratios under analysis (see Fig. \ref{fig:sfh1})}.}\\ 
\textcolor{black}{Fig. \ref{fig:sfh2} shows the time evolution of [O III]/H$\alpha$ and [Ne III]/[O II] for the three assumed SFHs, starting from the time of quenching.}
For the truncated SFH stopping at 200 Myr, the two emission line ratios drop by a factor $\sim$ 10 within $\sim$ 2 Myr and by more than a factor $\sim$ 1000 within $\sim$ 10 Myr from the quenching of the SF, similarly to the SSP case. \\In the case of an exponentially declining SFR, instead, they decrease following the decline of the SFR. In particular, we find that the star-forming region takes $\sim$ 80 Myr to become quiescent, reaching specific star-formation rate (sSFR) $\sim$ $10^{-11}$ yr$^{-1}$ (which are typical of quiescent galaxies) and, within this time interval, both [O III]/H$\alpha$ and [Ne III]/[O II] decrease by a factor $\sim$ 2.
\textcolor{black}{This decline corresponds to a decrease in log(U)$_{t}$ by only $\sim$ 0.2 dex, as illustrated in Fig. \ref{fig:plot_oiii_ha_logU}, implying that, when smoother SFHs are considered, log(U)$_t$ has a smoother decline.} Moreover, $\sim$ 500 Myr are necessary for the two emission line ratios to decline by a factor $\sim$10.\\ 
\textcolor{black}{It is interesting to note that the value of the two emission line ratios at the age at which the SF stops \textcolor{black}{(200 Myr for the truncated and $\sim$ 10 Myr for the exponentially declining SFH)} is lower by a factor $\sim$ \textcolor{black}{2} \textcolor{black}{for more complex SFHs than for a 0.01 Myr SSP with the same log(U)$_{0}=-3$}. Therefore, if an ionizing stellar population forms stars \textcolor{black}{continuously on a longer time interval}, its SED at the time of quenching is softer than the SED of a stellar population which forms all its stars into a single burst. \textcolor{black}{This can be due to the accumulation of long lived stars contributing mostly to the flux at longer wavelengths}. Moreover, the temporary increase of [O III]/H$\alpha$ and [Ne III]/[O II] due to Wolf-Rayet stars is less evident for more complex SFHs than for the SSP case, \textcolor{black}{since, even after the SF quenching, the evolution is dominated by the numerous generations of stars formed during time rather than by the short lived WR stars.% Note that, in order to better visualize and to compare the results for the SSP and the truncated case, we introduce the time $t_{\rm{quench}}$, which is the time at which the quenching begins, i.e. 0.01 Myr for the SSP and 200 Myr for the truncated SFH.
}}\\
\textcolor{black}{All these trends confirm that the decline of [O III]/H$\alpha$ and [Ne III]/[O II] takes place regardless of the shape of the assumed SFH and that very low values of these emission line ratios are expected whether galaxies have abruptly quenched their SF or they have gradually reached low levels of SF.} \\

%\begin{figure}%[t!]
  %  \resizebox{\hsize}{!}{\includegraphics{plot_oiii_ha_neiii_oii_evo_trunc_expo_200Myr_stb99_da7_compl2_2}}
 %   \caption{[O III]/H$\alpha$ and [Ne III]/[O II] emission line ratios as a function of time for different SFHs and Z = 0.02. Upper panels show the truncated (violet curve) and the exponentially declining (green curve) SFHs described in the text up to $\sim800$ Myr. Lower panels are a zoom on a time interval of $\sim10$ Myr from the SF quenching and compare the SSP and the truncated SFH (black curve) (the exponentially declining SFH in not shown since it has basically no variation in such a short time interval).  A logarithmic scale is adopted to highlight the rapid rise of ionizing photons characterizing the truncated and the exponentially declining SFHs at $t<t_{\rm{MS}}$(O stars). In the lower panels, $t_{\rm{quench}}$ is the time at which quenching begins (0.01 Myr for the SSP and 200 Myr for the truncated SFH).}
   %              \label{fig:sfh}
  %  \end{figure} 

\subsection{The time evolution of optical colours}
\label{sec:col}
\begin{figure}%[t!]
   % \resizebox{\hsize}{!}{\includegraphics{plot_u_r_r_i_stb99_vs_age_referee3_2giro}}
   %\resizebox{\hsize}{!}{\includegraphics{plot_cfr_u_r_SFHs}}
   \resizebox{\hsize}{!}{\includegraphics{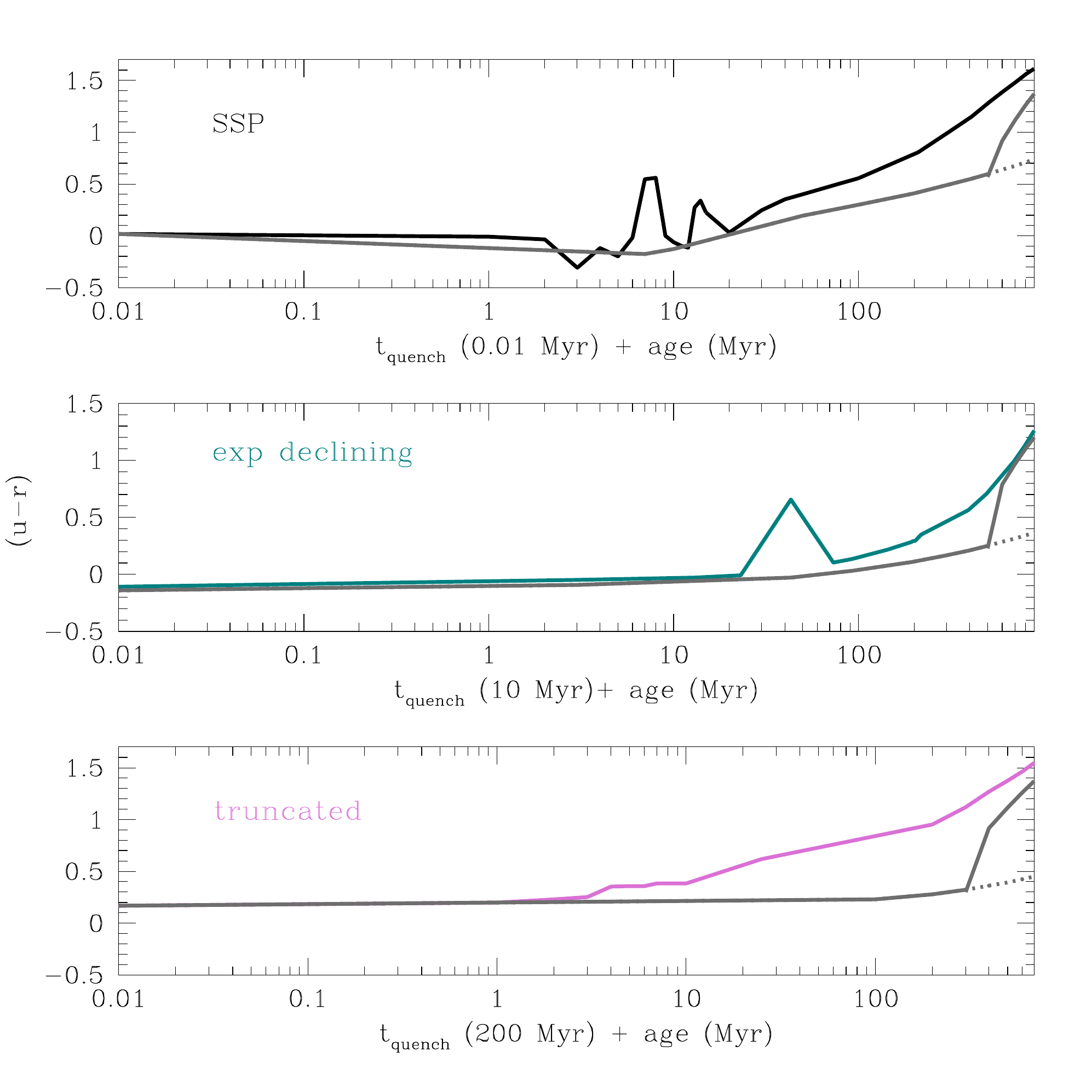}}
   
    \caption{\textcolor{black}{Time evolution of the optical $(u-r)$ colour, for solar metallicity and different SFHs. Black, violet and green curve refer to the SSP, truncated and exponentially declining SFHs considered in this work. Grey dotted and solid curves refer to a continuous SFH with SFR = 1 M$_{\sun}$ yr$^{-1}$ and  to a truncated SFR with SFR = 1 M$_{\sun}$ yr$^{-1}$ until 500 Myr and zero at later ages, respectively. From top to bottom, we show the colour evolution within 900 Myr from the quenching time of the SSP (i.e. 0.01 Myr), exponential ($\sim$ 10 Myr) and truncated ($\sim$ 200 Myr) SFHs considered in this work.}}
                 \label{fig:evo_colo}
    \end{figure}

\textcolor{black}{\textcolor{black}{%The softening of the SED associated to the SF quenching also implies an evolution of galaxy colours with time. 
%\textcolor{black}{%Differently from strong emission lines, whose presence in galaxy spectra is always related to star formation regardless of their complexity in terms of stellar content, galaxy colours are strongly affected by the presence of multiple populations inside a galaxy, and their interpretation is hampered by the differential reddening which they can produce. In this section, we show the evolution of galaxy colours as if they were isolated and not affected by the mutual effects of the several stellar populations forming the galaxy.}
The softening of the SED associated to the SF quenching also implies an evolution of galaxy colours with time. Since we are analyzing optical SDSS spectra, in this section we focus on the intrinsic $(u-r)$ optical colour, showing its evolution as a function of time in Fig. \ref{fig:evo_colo}. In particular, the three analyzed quenching SFHs are compared to a continuous SFH with SFR = 1 M$_{\sun}$ yr$^{-1}$ and to a quenching SFH with SFR = 1 M$_{\sun}$ yr$^{-1}$ until 500 Myr and zero at later ages. %The $(u-r)$ colour has a smoother increase in the truncated/exponentially declining case, where the various generations of stars formed during time dilute the effects related to short-lived stars.\\ 
The colour reddening is fast for the quenching SFHs, and colours are always redder than the colours of a continuous SF model even when a significant population of old stars has already been built up (i.e. after $\sim$ 500 Myr - 1 Gyr). %Moreover, for quenching SFHs, colours start to become considerably redder only after that the quenching has occurred.
We find that the typical $(u-r)$ colours of green valley ($1.5\lesssim(u-r)\lesssim2.5$, depending on mass) or red sequence ($(u-r)\gtrsim2-3$) galaxies (\citealp{Schawinski+2014}) are reached at later times (i.e. $\sim$ 1 Gyr) with respect to the time-scales considered so far (i.e. $\sim$ 10 -- 100 Myr). This happens even when an old stellar population is already present for a continuous SF model at the time of the SF halt (as in the case of the 500 Myr-truncated SFH).  It is important to note that all these trends refer to an individual quenching stellar population and could thus be diluted by the mutual effects of the multiple stellar populations inside a galaxy which are experiencing different evolutionary stages.}}

\subsection{The expected fractions of quenching candidates}
\label{sec:fq}

In the previous sections we found that [O III]/H$\alpha$ and [Ne III]/[O II] are very sensitive to the SF quenching and that star-forming galaxies characterized by low values of these two emission line ratios might be objects caught in the act of quenching. However, the precise time-scales that the emission line ratios are able to trace depend on the SFH of the central ionizing source. \textcolor{black}{Deriving these time-scales can be useful to understand what is the physical phenomenon associated to the SF quenching and to estimate the fraction of galaxies observed in the phase of quenching. In particular, the faster the SF is quenched, the faster the galaxies move from the blue cloud to the red sequence, and the less they clump in the transition zone.} To estimate the percentages of objects which are expected to be in the quenching phase, \textcolor{black}{we use the following eq. \ref{eq:fq}:
\begin{equation}
F_{QG}=t_{Q}/t_{DM}=t_{Q}\times sSFR~~,
\label{eq:fq}
\end{equation}
where $F_{QG}$ is the expected fraction of quenching galaxies, $t_{Q}$ is the time needed for the emission line ratios to decrease by a factor $\sim$ 10 (i.e. $\sim$ 10 Myr for the SSP/truncated case and $\sim$ 500 Myr for the exponentially declining SFH, see Sect. \ref{sec:que_diagn}) and $t_{DM}$ is the doubling-mass time (e.g. \citealp{Guzman+1997}, \citealp{Madau&Dickinson2014}), which we consider as a proxy of the typical life-time of a SF galaxy (\citealp{Guzman+1997} and \citealp{Greis+2016}). %If we consider the typical sSFR of local SF galaxies with 9 < log(M/M$_\sun$) < 12 (e.g. \citealp{Karim+2011}), this amounts to $\sim$ 3 ? 10 Gyr.
In particular, since the typical sSFRs of SF galaxies with redshift and masses comparable to ours (i.e. 0.2 < $z$ < 0.4 and 9 < log(M/M$_{\sun}$) < 12) range from $\sim$ 10$^{0.3}$  Gyr$^{-1}$ to $\sim$ 10$^{0.1}$  Gyr$^{-1}$ (e.g. \citealp{Karim+2011}), we can deduce that the typical $t_{DM}$ for our sample range from $\sim$ 3 Gyr to $\sim$ 10 Gyr, increasing for increasing mass.} \\
\textcolor{black}{The expected fractions of quenching galaxies are $\sim$ 0.06 $\%$ -- 0.2 $\%$ for the SSP/truncated SFHs and to $\sim$ 5 $\%$ -- 15 $\%$ for the exponentially declining one, respectively, decreasing for increasing mass.}
%We parametrize the lifetime of a SF galaxy as the time it needs for doubling its stellar mass (doubling mass time) given a SFR and a stellar mass (e.g. \citealp{Guzman+1997}, \citealp{Madau&Dickinson2014}). According to \citet{Guzman+1997} and \citet{Greis+2016}, for a constant SFR the doubling mass time is the inverse of the sSFR, and, if we consider the typical sSFR of local SF galaxies with $9 < $ log(M/M$_{\sun}$) < 12 (e.g. \citealp{Karim+2011}), this amounts to $\sim$ 3 -- 10 Gyr.
%Given this, if a sharp quenching is assumed, the fraction of quenching galaxies can range from $\sim$ 0.06 $\%$  to $\sim$ 0.2 $\%$, increasing with increasing mass. Instead, if an exponential decline of the star-formation is considered, the $\sim$ 500 Myr needed for the two emission line ratios to drop by more than a factor 10 leads to expected fractions of $\sim$ 5 -- 15 $\%$, depending on mass. 
\textcolor{black}{It is interesting to note that, once that a complete sample of quenching candidates is selected from a SF galaxy sample, the observed fractions $F_{QG}$ derived from data can help in disentangling what are the effective time-scales of the quenching mechanism at work. However, degeneracies (see Sect. \ref{sec:deg}) and  especially observational limits (e.g. the limiting fluxes of the analyzed emission line) can hamper the detections of the whole population of existing quenching candidates. Therefore, the derived percentages should be considered as lower limits, as well as the inferred quenching time-scales.}

  \section{Mitigating the ionization - metallicity degeneracy}  
  \label{sec:deg}

\begin{figure}%[t!]
     \resizebox{\hsize}{!}{\includegraphics{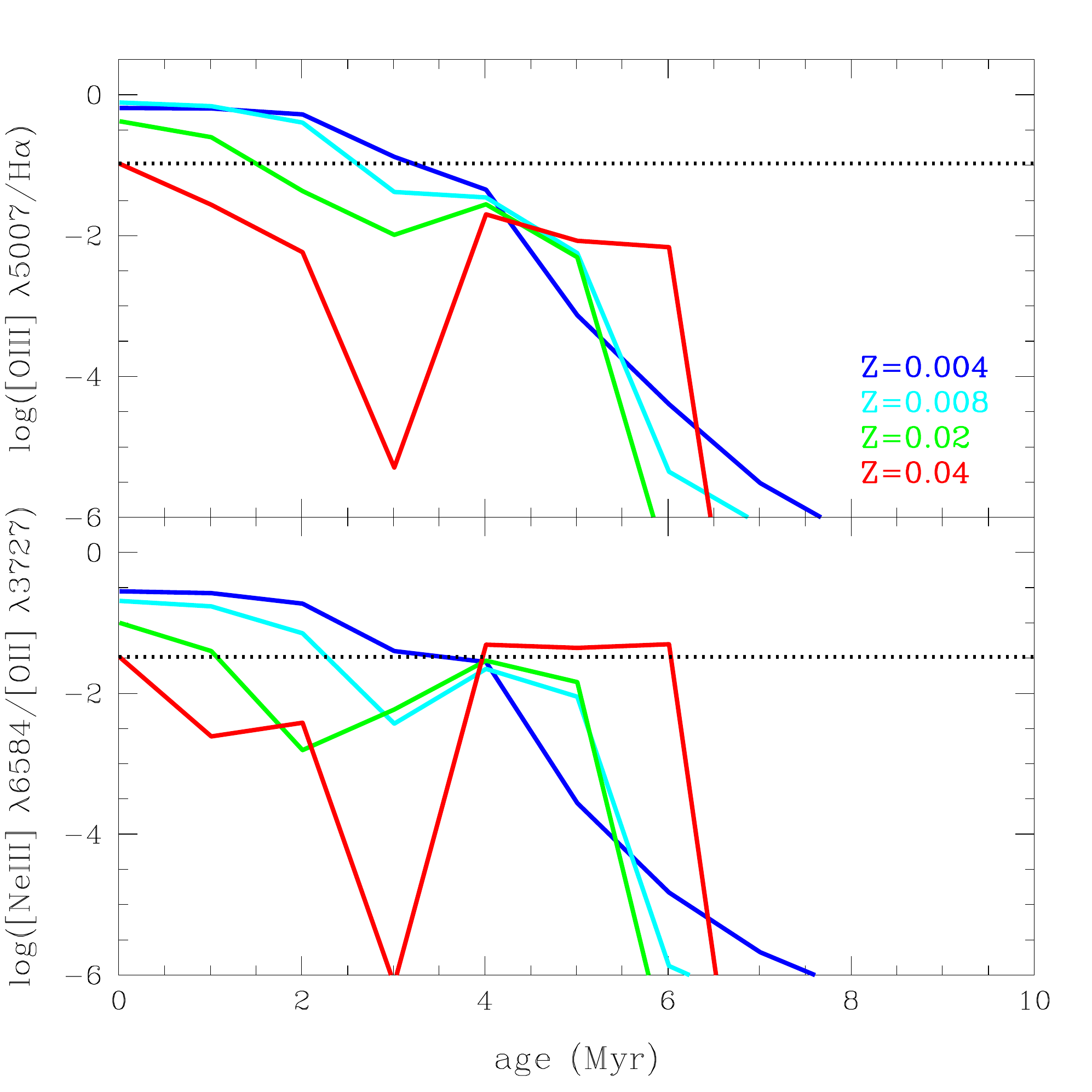}}
     \caption{[O III]/H$\alpha$ (top) and [Ne III]/[O II] (bottom) as a function of time, for log(U)$_{0}=-3$, and different metallicities (Z = 0.004, blue; Z = 0.008, cyan; Z = 0.02, green; Z = 0.04, red). In each panel, the black dotted line marks the initial value of the emission line ratios for Z = 0.04.}
     \label{fig:deg}
    \end{figure}

    \textcolor{black}{One of the limitations of our methodology is the metallicity degeneracy. Indeed, since higher metallicity ionizing SEDs are softer than lower metallicity ones due to the larger opacity of the ionizing stars (see Fig. \ref{fig:nphot}), they can have the same effects on the emitted spectrum and emission line ratios as quenched, low-ionization ones. Reducing this degeneracy is extremely important for our analysis, in order to exclude objects whose [O III]/H$\alpha$ and [Ne III]/[O II] are low because of high metallicities rather than low ionization.} The main effect of the metallicity degeneracy is illustrated in Fig. \ref{fig:deg}, for log(U)$_{0}=-3$. The same value of an emission line ratio can be produced by different combinations of age (i.e. log(U)$_{t}$) and metallicity. In particular, younger star-forming regions with higher metallicity can produce the same results as older star-forming regions with lower metallicity. This occurs for both [O III]/H$\alpha$ and [Ne III]/[O II].
    
\begin{figure}%[t!]
    \resizebox{\hsize}{!}{\includegraphics{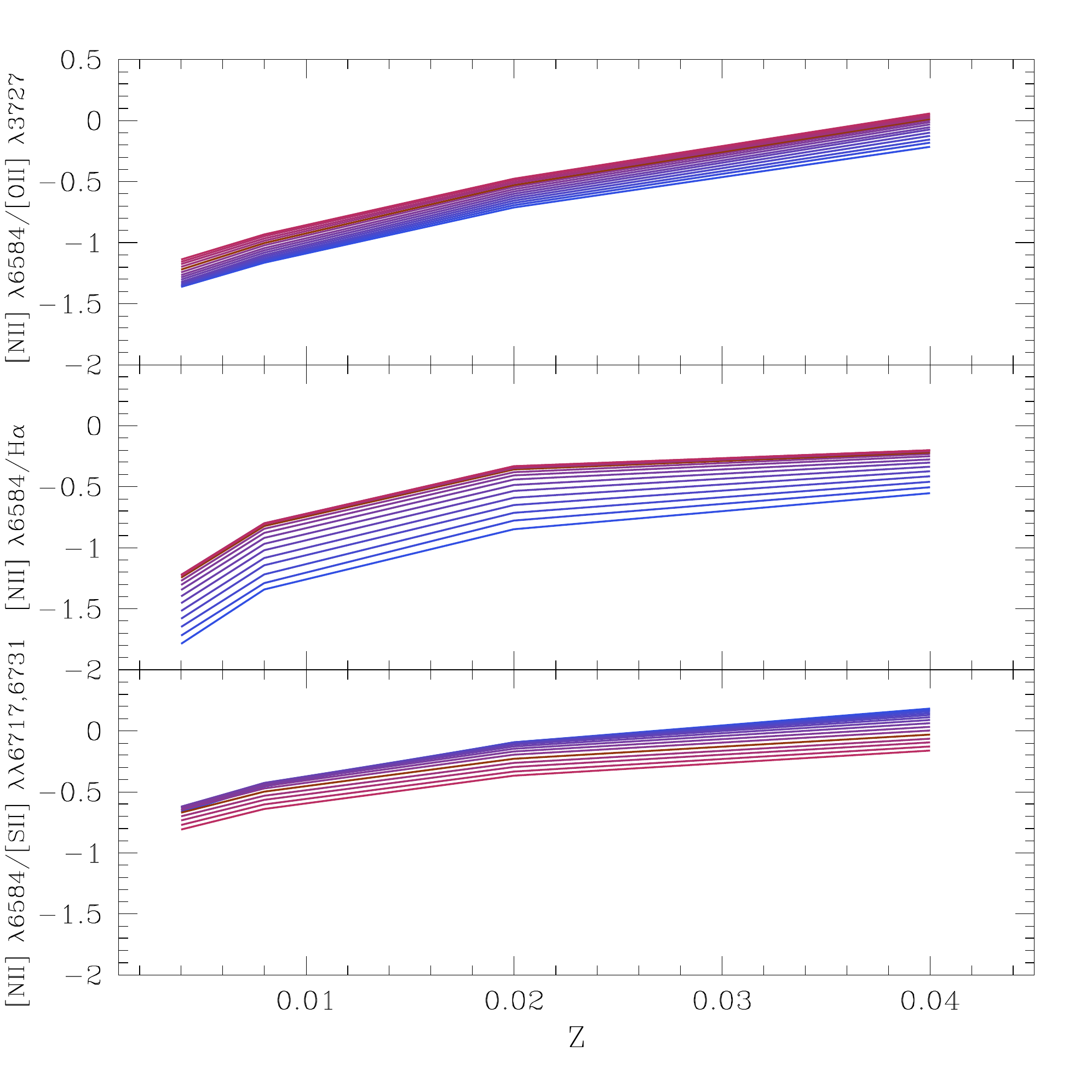}}
      \caption{Comparison between [N II]/[O II], [N II]/H$\alpha$ and [N II]/[S II] as a function of metallicity and log(U)$_{0}$. Colours are coded as in Fig. \ref{fig:oiii_Ha_age_met}.}
    \label{fig:nii_oii}
    \end{figure}

To mitigate the effects of the metallicity degeneracy, we suggest the use of independent pairs of emission line ratios, depending separately on ionization and metallicity as much as possible. In particular, we combine [O III]/H$\alpha$ and [Ne III]/[O II] with the [N II] $\lambda$6584/[O II] $\lambda$3727 ([N II]/[O II], hereafter) emission line ratio, whose behaviour in our model as a function of the stellar metallicity is illustrated in Fig. \ref{fig:nii_oii}, for a fixed age of 0.01 Myr. Here we also compare [N II]/[O II] with other two metallicity indicators, i.e. [N II]/H$\alpha$ (e.g. \citealp{Pettini&Pagel2004}; \citealp{Denicolo+2002}, \citealp{Nagao+2006}) and [N II] $\lambda$6584/[S II] $\lambda\lambda$6717,6731 (e.g. \citealp{Nagao+2006}) ([N II]/[S II], hereafter). While the [N II]/H$\alpha$ and [N II]/[S II] ratios flatten at high metallicities (\citealp{Kewley&Dopita2002}, \citealp{Levesque+2010}) and have a larger dependences on the ionization level, [N II]/[O II] scales smoothly from low to high abundances and has a little dependence on log(U)$_{0}$ (e.g. \citealp{vanZee+1998}, \citealp{Dopita+2000}, \citealp{Kewley&Dopita2002}). Moreover, even if the [N II]/[S II] ratio is less affected by reddening effects (it involves emission lines which are closer in wavelength), \textcolor{black}{it has a weaker dependence on metallicity with respect to} [N II]/[O II], which  has also the advantage of including stronger, and thus more easily detectable, emission lines. Therefore, it can be preferable once that the observed emission lines are corrected for dust extinction. In the following, in order to mitigate the metallicity degeneracy, we combine the two proposed quenching diagnostics with the [N II]/[O II] ratio.

\section{Identifying galaxies in the quenching phase}
\textcolor{black}{In the previous sections we found that \textcolor{black}{ galaxies which have quenched their SF in the recent past }can be characterized by low values of the [O III]/H$\alpha$ and the [Ne III]/[O II] emission line ratios. \textcolor{black}{In this section, we test the reliability of our approach by applying it to the SDSS sample described in Sect. \ref{sec:valid}, and analyzing the properties of some of the extreme candidates selected.}
 \textcolor{black}{However, it is important to remind that it is beyond the scope of this work to identify a complete sample of quenching candidates and to investigate their detailed properties, for which we refer to a companion paper (Quai et al. 2017, in prep).}\\
\textcolor{black}{Our methodology consists in selecting galaxies with high S/N H$\alpha$ in their spectra and lying at the lowest envelope (i.e. lowest log(U)$_0$) of the SF galaxy distribution within the [O III]/H$\alpha$ vs. [N II]/[O II] diagnostic diagram. Following the results discussed in Sect. \ref{sec:deg}, this plane is able to mitigate the metallicity degeneracy}}. \textcolor{black}{Since [O III]/H$\alpha$ and [N II]/[O II] are affected by dust extinction, we corrected the involved emission lines adopting the Calzetti extinction curve \citep{Calzetti+2000} and using the H$\alpha$/H$\beta$ ratio to estimate the nebular colour excess $\rm{E(B-V)}$. It is worth noting that, since the metallicity degeneracy strongly affects the [O III]/H$\beta$ vs. [N II]/H$\alpha$ plane, we do not use it to identify possible quenching candidates, although all the objects lying below log([O III]/H$\beta$)< -- 0.5 would fulfill our selection criterion (see Fig. \ref{fig:plot_griglia_definitiva1}).}

 \begin{figure}%[t!]
    \resizebox{\hsize}{!}{\includegraphics{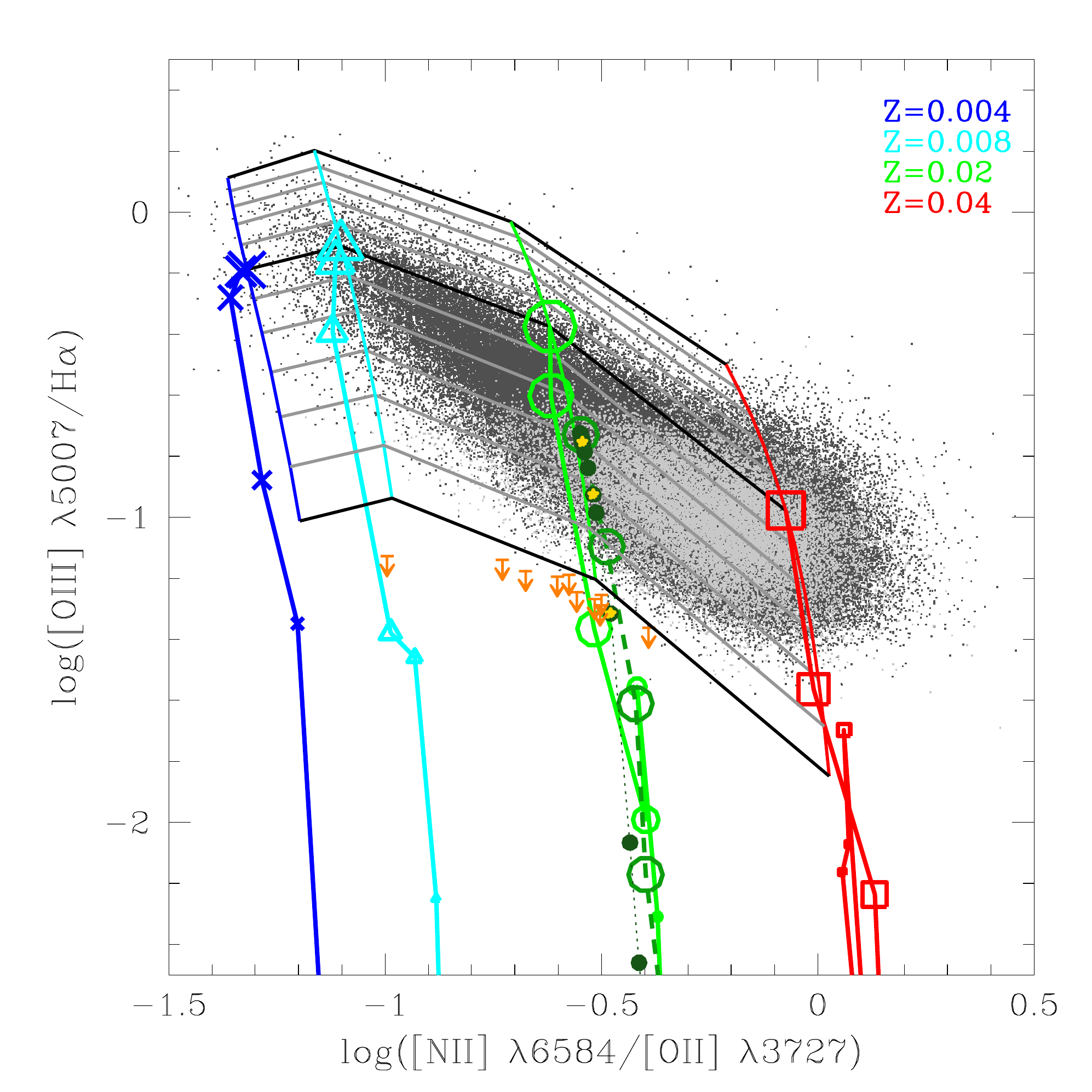}}
      \caption{Quenching candidates within the [O III]/H$\alpha$ vs. [N II]/[O II] plane. Dark grey points are galaxies extracted from the SDSS DR8 with S/N(H$\alpha$) > 5, S/N(H$\beta$) > 3 and S/N([N II]), S/N([O II]), S/N([O III]) > 2, while light grey points are galaxies with S/N([O III]) < 2. The superimposed grid is our set of fixed-age models with different metallicities, as in Fig. \ref{fig:plot_griglia_definitiva1}. Colored curves associated to different symbols are evolving-age models with an initial log(U)$_0= -3$ (with symbol size decreasing for increasing mass) for the four considered metallicities and with a time step of 1 Myr. For Z = 0.02, evolving-age models obtained for the truncated (dark green empty circles) and the exponentially declining (dark green filled circles) SFHs are shown with a time step of 1 Myr within the first 10 Myr after quenching, $\sim$ 20 Myr from 10 to 100 Myr after quenching, and 100 Myr even further. For the exponentially declining SFH, gold small stars mark the values of the emission line ratios corresponding to 10, 80, and 200 Myr after the SF quenching, from top to bottom. Orange downward arrows are the 10 extreme quenching candidates with S/N([O III]) < 2.}
    \label{fig:oii_Ha_nii_oii}
    \end{figure}
    
\begin{figure}%[t!]
    \resizebox{\hsize}{!}{\includegraphics{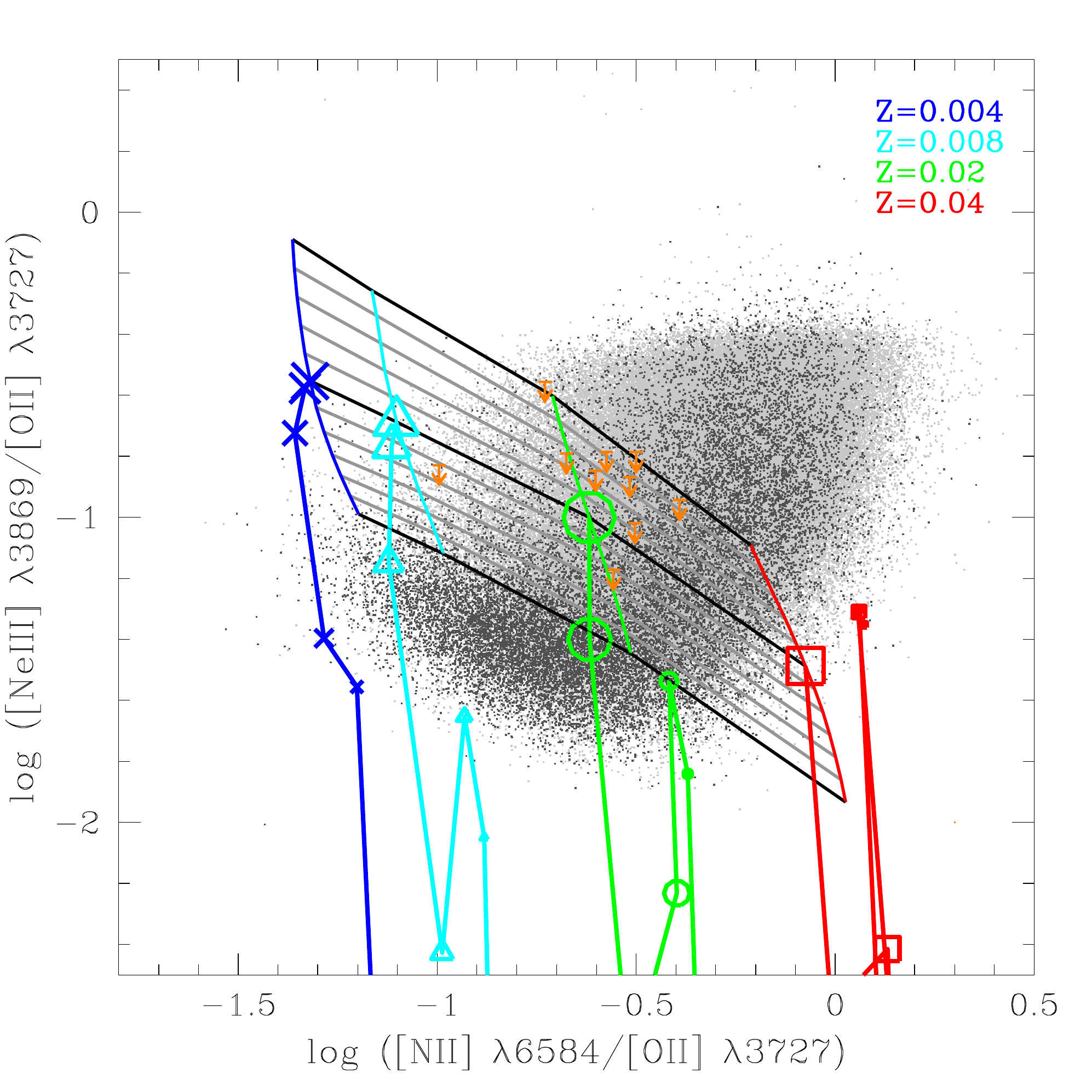}}
      \caption{Quenching candidates within the [Ne III]/[O II] vs. [N II]/[O II] plane. Dark grey points are galaxies extracted from the SDSS DR8 with S/N(H$\alpha$) > 5, S/N(H$\beta$) > 3, and S/N([N II]), S/N([O II]), S/N([Ne III]) > 2, while light grey points are galaxies with S/N([Ne III])< 2. Colours and symbols are defined as in Fig. \ref{fig:plot_griglia_definitiva1}. Orange downward arrows are the 10 extreme quenching candidates with S/N([O III]) < 2.
      }
    \label{fig:neiii_oii_nii_oii}
    \end{figure}

    As it is possible to note from Fig. \ref{fig:oii_Ha_nii_oii}, the bulk of the data distribution lies within a range of fixed-age ionization parameters going from log(U)$_0=$ -- 3 to log(U)$_0=$ -- 3.6, and spans the entire metallicity range of our models (the discrepancy between data and models at high values of [N II]/[O II] can be explained as in Sect. \ref{sec:valid}). The median of the data distribution can be described by $-3.2\lesssim$ log(U)$_{0}\lesssim-3$ decreasing for increasing metallicity, in agreement with the literature results concerning star-forming galaxies at low redshifts (e.g. \citealp{Dopita+2006}, \citealp{Nakajima&Ouchi2014}, \citealp{Shirazi+2014}, \citealp{Hayashi+2015}, \citealp{Kashino+2016},  \citealp{Onodera+2016}). Given this latter consideration, we also show evolving-age models characterized by log(U)$_0=-3$ at the time in which the SF stops, both in the SSP and the smoother SFHs cases (see Sect. \ref{sec:sfh}), for Z = 0.02. We find that the initial value of the [O III]/H$\alpha$ ratio is lower for smoother SFHs than for SSPs, as already discussed in Sect. \ref{sec:sfh}. Furthermore, for the exponentially declining SFH, the decrease of [O III]/H$\alpha$ is very slow, with the [O III]/H$\alpha$ values corresponding to the first 10 Myr after the SF quenching accumulating within $\lesssim$ 0.1 dex. \\ Since within this diagnostic diagram metallicity effects are separated from ionization ones, at each [N II]/[O II] (i.e. at each metallicity), objects with the lowest [O III]/H$\alpha$ ratios could be considered as quenching candidates. Starting from this, we select a subsample of 10 extreme objects with [O III]/H$\alpha$ below the values corresponding to the lowest log(U)$_0=-3.6$ and with S/N([O III]) < 2. \textcolor{black}{ \textcolor{black}{This ionization level is consistent with the quiescent phase of SSP evolving-age models, reached $\sim2$ Myr after quenching and corresponding to log(U)$_{t}\sim-3.2$. \textcolor{black}{If smoother SFHs are assumed, these very low levels of [O III]/H$\alpha$ are instead reached $\sim$ 200 Myr after the SF quenching, at log(U)$_t\sim-3.4$.}}}\\
  \textcolor{black}{
  The 10 selected objects are not detected in [O III] and they have upper limits $\lesssim$ -- 1.1 on the OIII/Halpha ratio. Moreover, basing on our models of [N II]/[O II] and on \citet{Tremonti+2004} models, they have intermediate metallicities (i.e. log([N II]/[O II]) ranging from $\sim$ -- 1 to $\sim$ -- 0.4), }
\textcolor{black}{and a faint [Ne III] emission line with S/N([Ne III]) < 2 (see Sect. \ref{sec:valid})}. This is compatible with our suggestion that galaxies that have quenched their SF in the recent past can be characterized by the absence of high-ionization lines. Moreover, it confirms that the two proposed emission line ratios are complementary in the identification of quenching candidates. \\
Fig. \ref{fig:neiii_oii_nii_oii}  illustrates the 10 objects in the [Ne III]/[O II] vs. [N II]/[O II] diagnostic diagram. Within this plane, the candidates have apparently high [Ne III]/[O II] ratios but, since [Ne III] is undetected, these values are actually upper limits, thus the real [Ne III]/[O II] ratios of the 10 objects can be much lower than the illustrated ones. The same holds for all the objects lying in the upper right or outside the model grid in Fig. \ref{fig:neiii_oii_nii_oii}, which are characterized by 2 < S/N([Ne III]) < 4. \\ 
%\textcolor{black}{We also verified that our method is not affected by the choice of the extinction curve adopted to correct SDSS data. In particular, we adopt the Allen (1976) extinction curve to make the comparison. We quantify the effect of the two extinction laws on [OIII]/H$\alpha$ and [NII]/[OII] calculating the median extinction differences ($A_{[NII]}-A_{[OII]}$) and ($A_{[OIII]}-A_{H\alpha}$) for both extinction curves.
%We find that  ($A_{[NII]}-A_{[OII]}$) is $\sim$ 0.35 in both cases, while ($A_{[OIII]}-A_{H\alpha}$) differs by only $\sim$ 0.03 mag, with the Allen (1976) extinction curve producing slightly higher [NII]/[OII] emission line ratios. This very small discrepancy lead us to conclude that our methodology  is not influenced by the choice of the extinction curve.}

%\textcolor{black}{In order to verify if the objects selected using our approach are compatible with being just-quenched objects, in the following we study some of their main properties. In this regard, it is important to note that the SFR derived from the H$\alpha$ emission line does not contain information on when the SF quenched in these objects. This is because we are catching them just after the SF halt, i.e. when a residual Ha emission can be provided by the colder O stars and by the B stars which are still present in the stellar population. For this reason, we better investigated their optical colors. Fig. \ref{fig:colormass} shows the $u-r$ vs. $r-i$ color-color diagram for our sample of SF galaxies and the 11 examples of quenching candidates. In particular corrected the two colors for dust extinction, assuming a \ref{Calzetti+2001} attenuation curve. Most of the 11 quenching galaxies have colors compatible with the ones predicted by our models in the case of an exponentially declining SFH after $\sim$ 100 Myr after the halt of the SF, which is the time needed for the two emission line ratios under analysis to disappear. This suggests that optical colors are influenced by the quenching of the SF within a relatively short time interval, getting redder, and that they can help to distinguish objects in the quenching phase. However, looking at the u-r -mass plane, the $u-r$ color of our 11 objects is bluer the the green valley typical colors, indicating that out objects have just entered the post-quenching phase.}

\subsection{\textcolor{black}{Properties of the quenching candidates}}
\begin{figure*}%[h!]
    %\resizebox{\hsize}{!}{\includegraphics{plot_11spec_oiiiND_extcorr_articolo1_referee}}
    %\resizebox{\hsize}{!}{\includegraphics{plot_11spec_oiiiND_referee_300317_cutout_crop}}
    \resizebox{\hsize}{!}{\includegraphics{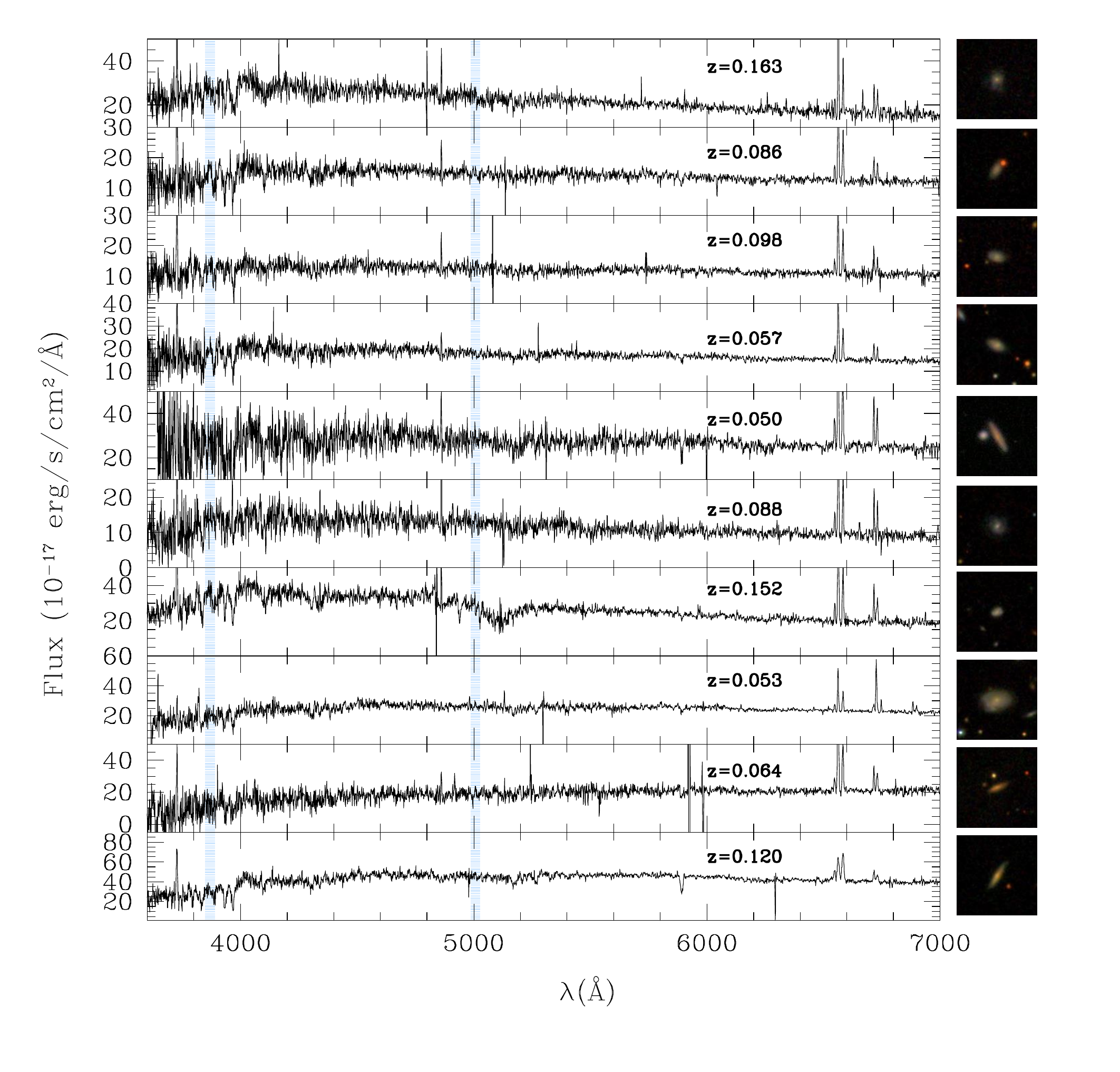}}
   % \resizebox{\hsize}{!}{\includegraphics{plot_11cand_crop_1}}
   
      \caption{Spectra of the 10 extreme quenching candidates corrected for dust exinction using the nebular colour excess $\rm{E(B-V)}$. The black curves are spectra corrected for dust extinction. \textcolor{black}{The emission lines discussed in this work are indicated on the top of the figure (from left to right: [O II], [Ne III], [O III], H$\alpha$, [N II], [S II]), and light blue shaded regions mark the not-detected [Ne III] and [O III] lines. Redshifts are reported for each object and morphologies are shown on the right side of each spectrum.}}
    \label{fig:morf1}
    \end{figure*}

\begin{figure*}%[h!]
    %\resizebox{\hsize}{!}{\includegraphics{stacked_oiiiND_ext_fino2_gli_lw1_6_referee}}
    %\resizebox{\hsize}{!}{\includegraphics{stacked_oiiiND_ext_tot}}
    \resizebox{\hsize}{!}{\includegraphics{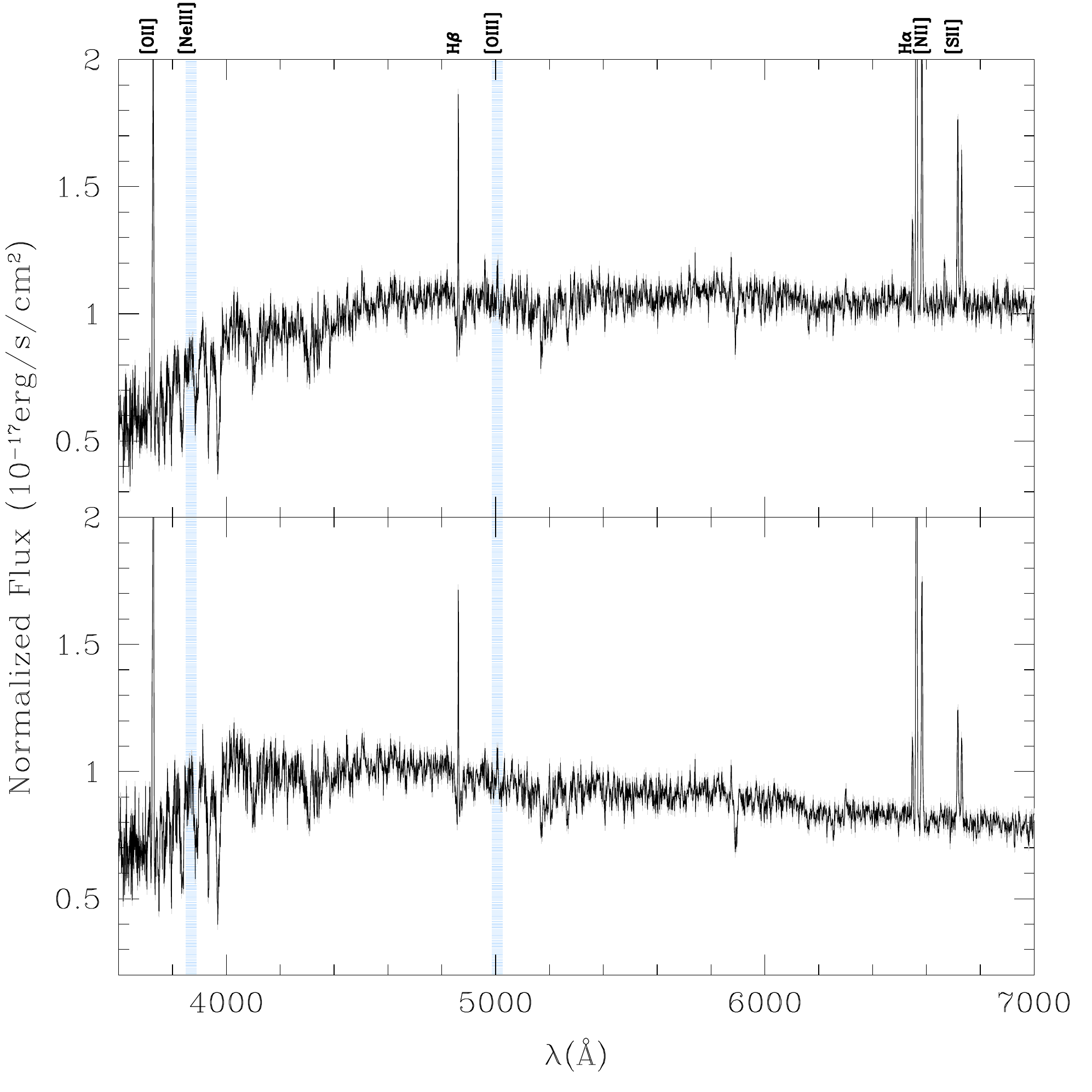}}
    \caption{\textcolor{black}{Median stacked spectrum of the 10 extreme quenching candidates. The top panel illustrates the median stacked spectrum (black curve) obtained from the dust-extincted quenching candidate spectra. The bottom panel shows the median stacked spectrum (black curve) obtained correcting the quenching candidates spectra for dust extinction, adopting the nebular $\rm{E(B-V)}$ \citep{Calzetti+2000} for both continuum and emission lines. Errors are shown in grey. \textcolor{black}{The emission lines discussed in this work are indicated on the top of the figure (from left to right: [O II], [Ne III], [O III], H$\alpha$, [N II], [S II]). Light blue shaded regions mark the not-detected [Ne III] and [O III] lines.}}}
    \label{fig:morf2}
    \end{figure*}

\begin{figure}%[t!]
    \resizebox{\hsize}{!}{\includegraphics{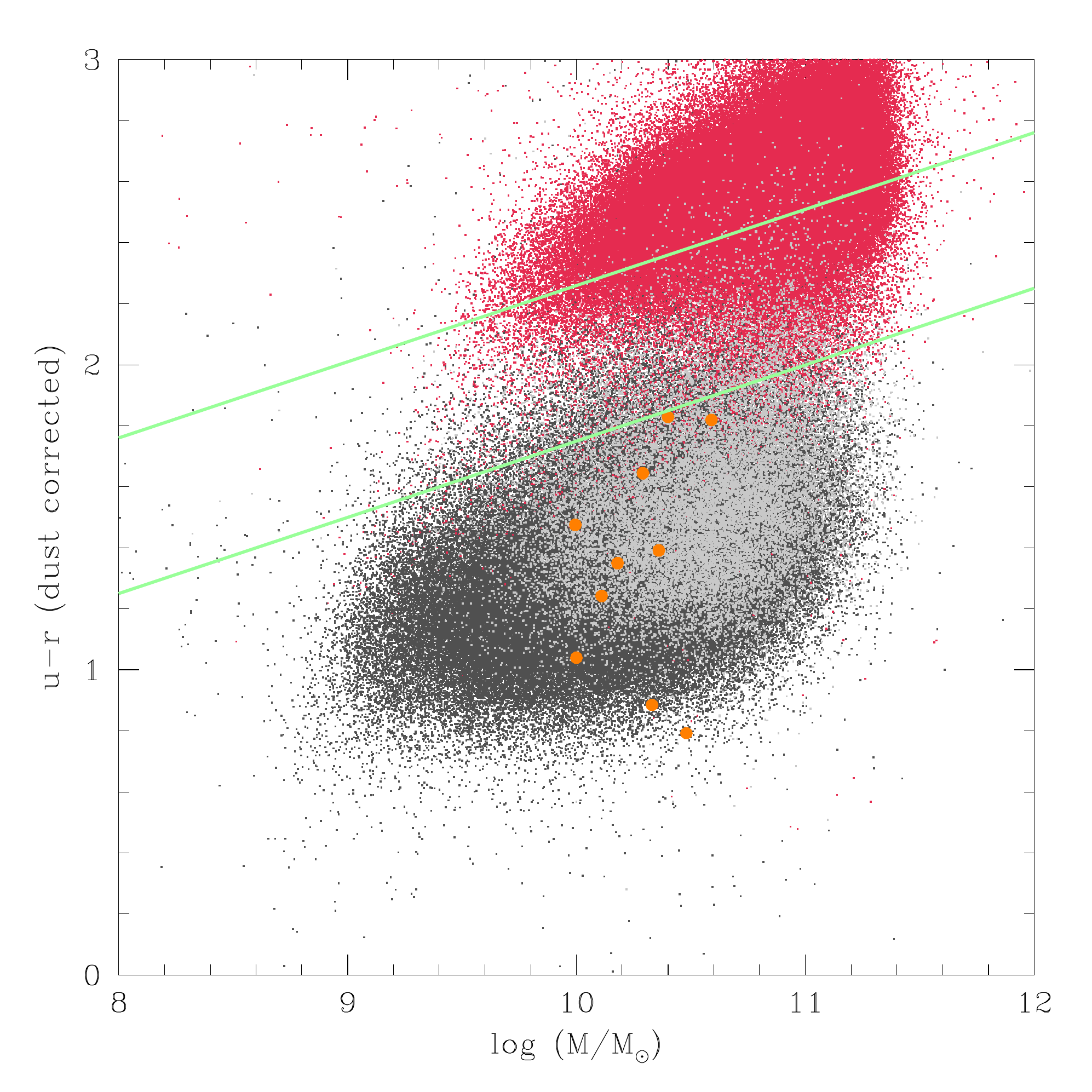}}
      \caption{\textcolor{black}{colour -- mass diagram for our galaxy sample ($(u-r)$ colours are corrected for dust extinction). Dark grey points are galaxies extracted from the SDSS DR8 with S/N(H$\alpha$) > 5, S/N(H$\beta$) > 3 and with S/N([N II]), S/N([O II]) and S/N([O III]) > 2, while light grey points are galaxies with S/N([O III]) < 2. Red points are galaxies with S/N (H$\alpha$) < 5 and/or EW(H$\alpha$) > 0 (see Quai et al. 2017, in prep. for further details). Orange circles are the 10 extreme quenching candidates with S/N([O III]) < 2. Green lines mark the green valley defined by \citet{Schawinski+2014}. This is taken as reference since it was derived from a sample of low-redshift galaxies, as ours.}}
    \label{fig:colormass}
    \end{figure}

\textcolor{black}{In this section we \textcolor{black}{illustrate} the main properties of the 10 selected quenching galaxies, verifying if they are compatible with a recent quenching of the SF.}\\
 %Note that this extreme population can be identified also within the BPT diagram, at [O III]/H$\beta\lesssim$ 0.5 (see Sect. \ref{sec:valid}), but in this plane, as already mentioned (see Sect. \ref{sec:valid}), we are strongly affected by the metallicity degeneracy.}\\
 Figs. \ref{fig:morf1} and \ref{fig:morf2} illustrate the candidate spectra, corrected for dust extinction adopting the Calzetti extinction curve (\citealp{Calzetti+2000}) and the H$\alpha$/H$\beta$ ratio to estimate the nebular colour excess $\rm{E(B-V)}$. \textcolor{black}{In particular, we applied the same correction for both the continuum and the emission lines. %their dust-corrected SEDs are substantially blue. \textcolor{red}{This suggests that they could have experienced the SF quenching in the very recent past,} \textcolor{red}{and thus their optical SEDs have not had time to \textcolor{black}{age and get significantly redder}}. \textcolor{black}{ in agreement with the optical colors behaviour.} This is indeed what we would expect from our diagnostics to select galaxies where the SF has been just interrupted. 
Spectra lack of strong absorption Balmer lines, which are characteristic of the post-starburst phase. This indicates that the candidates may have been caught in an earlier evolutionary phase with respect to the post-starburst (E+A, K+A) one (e.g. \citealp{Dressler&Gunn1983}, \citealp{Couch&Sharples1987}, \citealp{Poggianti+2008}). Moreover, they have strong H$\alpha$ and [O II] lines\textcolor{black}{,  in agreement with our predictions that low-ionization lines decrease by a small factor with respect to high-ionization ones in just quenched objects (cfr Sect. \ref{sec:que_diagn}). 
%since, if quenching has occurred in the very recent past, the majority of stars (i.e. colder O stars and B stars) able to produce these low-ionization lines are not yet disappeared. This implies that the H$\alpha$ emission in just-quenched objects is an indicator of the residual SF provided by less massive stars, since the galaxy SF is actually ended 10 -- 80 Myr ago, basing on our models.-0.1475796243 0.5595583477
}}
Fig. \ref{fig:morf2} \textcolor{black}{shows the median stacked spectrum of the 10 candidates obtained by stacking \textcolor{black}{the individual spectra before and after applying the correction for dust extinction. In this latter case we assumed the \citet{Calzetti+2000} extinction curve and the same nebular colour excess $\rm{E(B-V)}$ for both emission lines and continuum.} \textcolor{black}{The 10 candidates have $0.2 \lesssim$ $\rm{E(B-V)}$ $\lesssim0.8$, with a mean $\rm{E(B-V)}$ $\sim0.5 \pm 0.2$. The large $\rm{E(B-V)}$ of some of them explain why the stacked spectrum derived from individual spectra corrected for dust extinction is significantly bluer than the one derived from dust-extincted spectra.} We find that, even after the stacking, which increases the S/N of the individual spectra, \textcolor{black}{the stacked spectrum has very faint high-ionization lines} in both cases. This confirms that the absence of the high-ionization emission lines \textcolor{black}{is a real feature} and not an artifact due to low signal in the individual spectra}.\\
\textcolor{black}{The 10 candidates have high H$\alpha$ luminosities L(H$\alpha$) (as expected from the evolution described in Sect. \ref{sec:que_diagn}), ranging from 10$^{40}$ to 10$^{42}$ erg/s/\AA , and a median SFR $\sim1.4\pm1.7~M_{\sun}~yr^{-1}$. However, we remind that the usual recipes to convert L(H$\alpha$) into SFR cannot be applied to our galaxies because most O stars are missing compared to a galaxy where star-formation is still ongoing. Therefore, the L(H$\alpha$)-derived SFRs represent the past SFRs (i.e. the ones before the quenching) and thus should be considered as upper limits.} \textcolor{black}{We also find that the 10 candidates have Elliptical and S0 morphologies. We refer to the companion paper Quai et al. (2017, in prep.) for a detailed discussion about the typical morphologies of complete samples of quenching candidates and how they relate to the parent population of SF galaxies.}\\
%\textcolor{black}{This implies that the H$\alpha$ emission in just-quenched objects is an indicator of the just-halted SF (which actually ended 10 -- 80 Myr before). Therefore, we expect the H$\alpha$ luminosities of these object to be high in these source and the SFR derived from the H$\alpha$ luminosity L(H$\alpha$) to be just an upper limit.}
\textcolor{black}{The nature of the 10 quenching candidates is also shown in Fig. \ref{fig:colormass}, which illustrates the $(u-r)$ -- mass diagram for our sample. Following \citet{Schawinski+2014}, we corrected colours for dust extinction assuming the colour excess of the stellar continuum, defined as $\rm{E_{s}(B-V)}=0.44\times\rm{E(B-V)}$, where $\rm{E(B-V)}$ is the nebular continuum adopted so far (\citealp{Calzetti+2000}). Colours are compatible with the evolution discussed in Sect. \ref{sec:col} at early times after the SF halt, and are bluer than the green valley $(u-r)$ colours (corrected for dust extinction) derived by \citet{Schawinski+2014}. This suggests that the 10 galaxies may be at the beginning of the quenching process, not being yet as red as green valley galaxies. Furthermore, it is interesting to note that they have log(M/M$_{\sun}$) $\sim$ 10, compatible with the mass of the passive galaxy population building up at low-intermediate redshifts (e.g. \citealp{Pozzetti+2010})}.\\

\begin{figure}%[t!]
    \resizebox{\hsize}{!}{\includegraphics{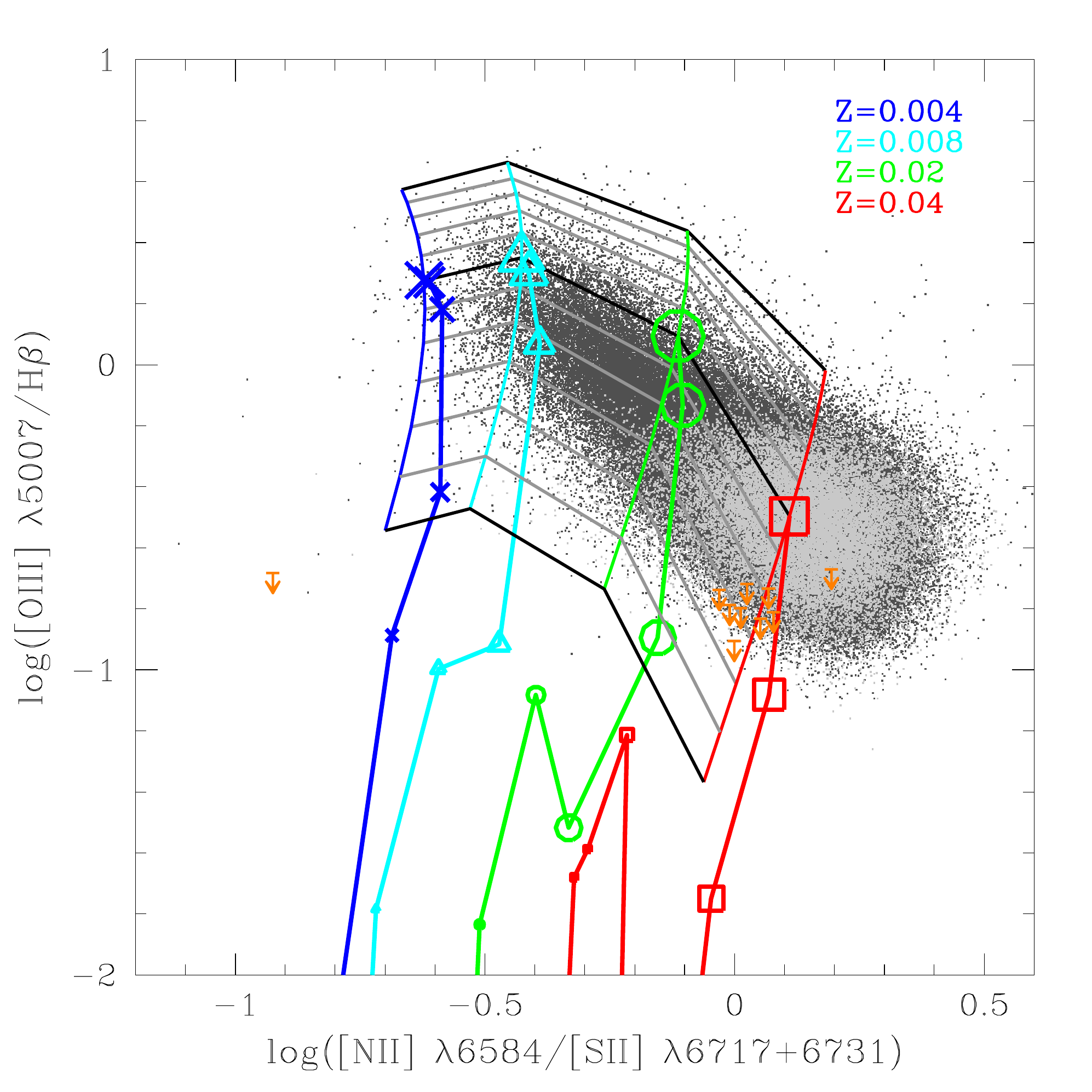}}
      \caption{\textcolor{black}{Quenching candidates within the [O III]/H$\beta$ vs. [N II]/[S II] plane. Dark grey points are galaxies extracted from the SDSS DR8 with S/N(H$\beta$) > 3 and S/N([N II]), S/N([S II]), S/N([O III]) > 2, while light grey points are galaxies with S/N([O III]) < 2. Colours and symbols of the grid are defined as in Fig. \ref{fig:plot_griglia_definitiva1}. Orange arrows are the 10 extreme quenching candidates with S/N([O III]) < 2.}
      }
    \label{fig:oiii_hb_nii_sii}
   \end{figure}

\textcolor{black}{
Since the  [O III]/H$\alpha$ and the [N II]/[O II] emission line ratios are affected by dust extinction, we verified if our results are influenced 
\textcolor{black}{by the choice of the extinction law assumed to correct our data.}\textcolor{black}{ We adopt the \citet{Allen1976} extinction curve, comparing the results with the ones obtained so far using the \citet{Calzetti+2000} one. We quantify the effect of the two extinction laws on the emission line ratios [O III]/H$\alpha$ and [N II]/[O II] by calculating the median extinction differences ($A_{[N II]}-A_{[O II]}$) and ($A_{[O III]}-A_{H\alpha}$) for both cases.
We find that ($A_{[O III]}-A_{H\alpha}$) is $\sim$ 0.35 in both cases, while ($A_{[N II]}-A_{[O II]}$) differs by $\sim$ 0.03 mag, with the \citet{Allen1976} extinction curve producing slightly higher [N II]/[O II]. This very small discrepancy leads us to conclude that our approach is not influenced by the choice of the extinction curve.}}\\
    We also investigated if our results are valid when dust-free diagnostic diagrams are used, which would have the advantage of producing results not relying on the assumption of an extinction law.
    Fig. \ref {fig:oiii_hb_nii_sii} shows the 10 quenching candidates within the [O III]/H$\beta$ vs. [N II]/[S II] plane, which involves ratios between emission lines closer in wavelength, and thus is basically unaffected by dust extinction. Although the discrepancy between models and data at the highest metallicities is still present, within this plane the 10 galaxies are located at the lowest edge of the data distribution, suggesting that the [O III]/H$\beta$ can be used as a dust-free, quenching diagnostic, similarly to [O III]/H$\alpha$. However, the [N II]/[S II] ratio, although being appropriate to avoid the uncertainties related to dust extinction, involves weaker lines like [S II] and is less efficient in separating different metallicities compared to [N II]/[O II], also producing larger errors relative to the covered range in metallicity. \\
%\textcolor{black}{Since the  [O III]/H$\alpha$ and the [N II]/[O II] emission line ratios are affected by dust extinction, it can be useful to verify if our approach is valid also when dust-free diagnostic diagrams are used, since they would have the advantage of producing results not relying on the assumption of an extinction low. Fig. \ref {fig:oiii_hb_nii_sii} shows the 11 quenching candidates within the [O III]/H$\beta$ vs. [N II]/[S II] plane, which involves ratios between emission lines closer in wavelength, and thus is basically unaffected by dust extinction. Although the discrepancy between models and data at the highest metallicities is still present, within this plane the 11 galaxies are located at the lowest edge of the data distribution, suggesting that the [O III]/H$\beta$ can be used as a dust-free, quenching diagnostic, similarly to [O III]/H$\alpha$. However, the [N II]/[S II] ratio, although being appropriate to avoid the uncertainties related to dust extinction, involves weaker lines like [S II] and is less efficient in separating different metallicities compared to [N II]/[O II], also producing larger errors relative to the covered range in metallicity. 

\textcolor{black}{Since the 10 selected objects are just an example of quenching candidates, they do not represent a complete sample of just quenched galaxies and cannot be used to derive reliable quenching candidate fractions (see Sect. \ref{sec:fq}). We refer to Quai et al (2017, in prep.) for a detailed discussion about them.}
%   However, being basically the extreme tail of the [O III] not-detected subsample of $\sim26,000$ objects, if we consider this latter as the whole quenching population with respect to the total sample of 174,000 galaxies, we obtain $F_{q} \sim 15 \%$, compatible with the predicted value for a smooth SF quenching (see Sect. \ref{sec:fq}).}

\section{Conclusions}
In this work, we propose a \textcolor{black}{methodology} aimed at identifying galaxies in the phase which immediately follows the quenching of the star formation, 
\textcolor{black}{under the assumption of a fast quenching (shorter than 200 Myr), both for an extreme sharp case and for a more realistic, smoother SF decline.}
%\textcolor{black}{under the main assumption that galaxies have experienced a sharp quenching of the SF (e.g. \citealp{Schawinski+2014}).} 
Our approach is based on the use of specific emission line ratios involving high- and low-ionization lines, which are expected to be strongly affected by quenching, rapidly dropping when the most massive O stars able to produce high-ionization lines, disappear. Our main findings can be summarized as follows.

\begin{enumerate}

\item We focus in particular on the [O III]/H$\alpha$ and [Ne III]/[O II] emission line ratios, modelling them by means of the CLOUDY \citep{Ferland+2013} photoionization code. We find that they are able to trace the phase just after the SF quenching, decreasing even by more than a factor 10 after its occurrence. The time-scales of this decline depend on the SFH adopted to characterize the ionizing stellar population. In particular, if \textcolor{black}{the extreme case of} a sharp quenching is assumed, the emission line ratios definitively drop on time-scales of $\sim$ 10 Myr after the SF quenching, while, if a smoother \textcolor{black}{and more realistic} quenching is considered (e.g. an exponentially declining SFR with an $e$-folding time $\tau$ = 200 Myr) they decline by a factor $\sim$ 2 within the time interval needed by the star-forming region to become quiescent (i.e. $\sim$ 80 Myr);\\

\item Our methodology is valid even when different synthetic stellar spectra are assumed to characterize the central ionizing source. In particular, in the case of a sharp quenching, we find a good agreement between Starburst99 and BC03 models: the decline of the two emission line ratios occurs in both cases and happens on similar time-scales, at each metallicity. This is an interesting result, since BC03 models are generally used to describe more advanced phases of galaxy evolution;\\

\item We find that our approach is influenced by the ionization -- metallicity degeneracy, since more metallic ionizing sources, being characterized by a low number of hard UV ionizing photons, can produce the same low values of the analysed emission line ratios as quenching ionizing sources. We find that this degeneracy can be reduced using pairs of independent emission line ratios, separately related to metallicity and ionization. In particular, due to its strong dependence on metallicity and weak dependence on ionization parameter, we propose the [N II]/[O II] ratio as metallicity diagnostic to mitigate the degeneracy, and alternatively the [N II]/[S II] one, less affected by dust attenuation but involving weaker lines like [S II];\\

%\item \textcolor{black}{We find that the $\rm{n_{H}}$  degeneracy has, in principle, an impact on the theoretical trends discussed, especially when high $\rm{n_{H}}$  values are assumed. However, our results are not significantly affected by this issue, since the values of $\rm{n_{H}}$ at which the degeneracy is stronger are well above those of typical H II regions (i.e. $\rm{n_{H}}\sim10^{2}cm^{-3}$), excluding that the very low values of [O III]/H$\alpha$ and [Ne III]/[O II] found in this work are due to density effects rather than quenching effects;}\\
\item We compare our models to a sample of $\sim$ 174,000 SDSS DR8 star-forming galaxies, in order to verify if objects characterized by low values of [O III]/H$\alpha$ and [Ne III]/[O II] are indeed present in the global galaxy population and if their properties are compatible with a recent SF quenching. We mainly use the [O III]/H$\alpha$ vs. [N II]/[O II] diagnostic diagram, since it is able to separate metallicity effects from ionization ones.
Within this plane, we identify 10 objects with extremely low [O III]/H$\alpha$ ratios and S/N([O III]) < 2. \textcolor{black}{We find that their spectra are characterized by the absence of the [Ne III] line but rather strong H$\alpha$ and [O II]. \textcolor{black}{All these properties agree with the hypothesis that they may have quenched their SF in the recent past;}}\\

\item \textcolor{black}{Within the dust-corrected $(u-r)$ -- mass diagram, the 10 candidates lie outside the green valley, due to the bluer colours. These latter are compatible with the evolution predicted by our models after $\sim$ 10 -- 100 Myr from the SF quenching, further suggesting that the 10 objects may be at the beginning of the quenching process;}\\%However, it is important to note that the predicted colour trends refers to an individual quenching stellar population and could thus be diluted if all the stellar populations inside a galaxy which are experiencing different evolutionary stages are taken into account;}\\

%blue dust-corrected spectra and strong H$\alpha$ and [O II] emission lines. \textcolor{red}{These properties suggest that they may have quenched their SF in the recent past. %their optical SEDs have not had time to get older and redder and their low-ionization emission lines are not yet disappeared.

 \item We find that \textcolor{black}{our methodology is independent from the dust extinction curve adopted to correct the data. Moreover,} the [O III]/H$\beta$ ratio is similar to [O III]/H$\alpha$ in identifying quenching candidates, while [N II]/[S II], although being unaffected by dust extinction, has the disadvantage to involve the weaker [S II] line and of being less efficient in separating metallicities with respect to [N II]/[O II], also producing larger errors relative to the covered range in metallicity;\\

\item We verify that our approach is valid also when dust-free diagnostic diagrams, such as the [O III]/H$\beta$ vs. [N II]/[S II] one, are used. Indeed, the 10 quenching candidates lie at the lowest edge of the data distribution also in this plane. %Moreover, taken at face value, the 11 galaxies would imply a quenching time-scale of $\sim$ 10 -- 80 Myr. %This will be investigated in detail in a companion paper by Quai et al. 2017, in prep.

\end{enumerate}

\textcolor{black}{The proposed methodology suggests that emission line ratios involving high- and low-ionization lines are powerful tools to identify galaxies in the quenching phase. Moreover, this kind of approach has the advantage of being applicable also at higher redshifts, once that suitable pairs of emission lines are chosen. Therefore, our methodology can be used for the analysis of large spectroscopic surveys data such as JWST \citep{Gardner+2006}, WFIRST \citep{Spergel+2013} and Euclid \citep{Laureijs+2011}, which identify a large number of H$\alpha$ emitters at $z>1$ (e.g. \citealp{Pozzetti+2016}). \textcolor{black}{Since the 10 candidates identified in this work are just examples of quenching galaxies, they cannot be used to derive reliable quenching fractions. However,} our results indicate that, once that a complete sample of quenching candidates is identified within a SF galaxy sample, their observed fraction, compared to the global population of star-forming galaxies, could be used to disentangle what is the quenching mechanism at work and what are its typical time-scales. We refer to a companion paper (Quai et al., 2017, in prep.) both for this purpose and for the investigation of the detailed properties of \textcolor{black}{complete samples} of quenching candidates.}

\section*{Acknowledgements}

The authors are grateful to Emanuele Daddi and Gra\.zyna Stasi\'nska for useful discussion.
ACit is also grateful to Lucia Armillotta, Alfonso Veropalumbo and Gianni Zamorani for helpful discussion and suggestions.
The authors also acknowledge the support of the grant PRIN MIUR 2010-2011 {\em The 
dark Universe and the cosmic evolution of baryons: from current surveys to Euclid}, PRIN MIUR 2015 {\em Cosmology and Fundamental Physics: illuminating the Dark Universe with Euclid} and the grants ASI n.I/023/12/0 {\em Attività relative alla fase B2/C per la missione Euclid}.
Funding for the SDSS and SDSS-II has been provided by the Alfred P. Sloan Foundation, the Participating Institutions, the National Science Foundation, the U.S. Department of Energy, the National Aeronautics and Space Administration, the Japanese Monbukagakusho, the Max Planck Society, and the Higher Education Funding Council for England. The SDSS Web Site is http://www.sdss.org/. The SDSS is managed by the Astrophysical Research Consortium for the Participating Institutions. The Participating Institutions are the American Museum of Natural History, Astrophysical Institute Potsdam, University of Basel, University of Cambridge, Case Western Reserve University, University of Chicago, Drexel University, Fermilab, the Institute for Advanced Study, the Japan Participation Group, Johns Hopkins University, the Joint Institute for Nuclear Astrophysics, the Kavli Institute for Particle Astrophysics and Cosmology, the Korean Scientist Group, the Chinese Academy of Sciences (LAMOST), Los Alamos National Laboratory, the Max-Planck-Institute for Astronomy (MPIA), the Max-Planck-Institute for Astrophysics (MPA), New Mexico State University, Ohio State University, University of Pittsburgh, University of Portsmouth, Princeton University, the United States Naval Observatory, and the University of Washington.

%%%%%%%%%%%%%%%%%%%%%%%%%%%%%%%%%%%%%%%%%%%%%%%%%%



\bibliographystyle{mnras}
\bibliography{art_cloudy} % if your bibtex file is called example.bib

%\begin{thebibliography}{99}
%\end{thebibliography}

%%%%%%%%%%%%%%%%%%%%%%%%%%%%%%%%%%%%%%%%%%%%%%%%%%

%%%%%%%%%%%%%%%%% APPENDICES %%%%%%%%%%%%%%%%%%%%%

%\appendix



%%%%%%%%%%%%%%%%%%%%%%%%%%%%%%%%%%%%%%%%%%%%%%%%%%

% Don't change these lines
\bsp	% typesetting comment
\label{lastpage}
\end{document}